# Chapter 13

## Probing the formation of the seeds of supermassive black holes with gravitational waves[1]

Monica Colpi[1,2]

[1] *Department of Physics G. Occhialini, University of Milano Bicocca*
[2] *National Institute of Nuclear Physics, Milano Bicocca*
*Piazza della Scienza 3, I 20123 Milano, Italy*
*monica.colpi@unimib.it*

The existence of black holes in the intermediate-mass interval between $10^2\,\mathrm{M}_\odot$ and $10^5\,\mathrm{M}_\odot$, filling the *gap* between the stellar and the supermassive black holes is a key prediction to explain the origin of luminous QSOs at redshifts $z$ as large as 7. There is a sheer difficulty in forming a $10^9\,\mathrm{M}_\odot$ black hole in less than one billion years. This has led to the concept of *seed black holes*. They are high-$z$ intermediate-mass black holes that formed during the dark ages of the universe. Seeds are a transient population, which later grew massive through episodes of accretion and mergers. In this chapter we explore the possibility of discovering seed black holes and track their growth across all cosmic epoch, by detecting the gravitational wave signal they emit at the time of their coalescence, if/when they pair to form close binaries. We show that the ESA's LISA mission for the detection of low-frequency gravitational waves will be paramount in granting this insight. Gravitational waves travel unimpeded through the cosmos and carry exquisite information on the masses and spins of the merging black holes. To this purpose we introduce key concepts on the gravitational wave emission from binaries, describing briefly their formation pathway during halo-halo mergers and galaxy collisions.

[1] Preprint of a review volume chapter to be published in Latif, M., & Schleicher, D.R.G., "Probing the formation of the seeds of supermassive black holes with gravitational waves", Formation of the First Black Holes, 2018 ©Copyright World Scientific Publishing Company, https://www.worldscientific.com/worldscibooks/10.1142/10652

## 1. Paving the way to unveil black holes in the high redshift universe

The simplest object in Nature is a *black hole* (BH)* exact solution of Einstein's field equations representing a spinning, point mass $M$, in otherwise empty space. BHs are believed to power many astrophysical sources, from the luminous quasi-stellar objects (QSOs) to the galactic X-ray sources, through the conversion of gravitational energy into electromagnetic (EM) radiation. In this chapter we show that BHs in close binaries are among the most powerful sources of gravitational waves (GWs) in the universe.

EM observations reveal the occurrence of BHs in two different flavors: the *stellar-origin black holes* (sBHs) and the *supermassive black holes* (SMBHs). sBHs are widespread in all the galaxies of the universe, as they are the relic of massive stars. They started to form in pristine, dark matter halos as early as redshift $z \sim 20$ from the collapse of massive population III stars (Madau and Rees, 2001), and continue to form in the galaxies as stars do. When a sBH is bound to a star in a close binary, it lights up as X-ray source, but when bound to a sBH, the system emits GWs. The GW signal is initially weak, but it grows loud when the two sBHs are in the process of merging to form a new sBH. GW150914 (Abbott *et al.*, 2016a), the first cosmic source of GWs hosting two coalescing sBHs, is a chief example.

Prior to the discovery of GW150914, the measurements of sBH masses in the Milky Way Galaxy were indicating a narrow distribution around $7.8 \pm 1.2\,\mathrm{M}_\odot$ (Özel *et al.*, 2010). The discovery of "heavy" sBHs, with mass up to $\sim 36^{+5}_{-4}\,\mathrm{M}_\odot$, in GW150914, has shown that the sBH mass spectrum is wider than EM observations revealed so far. With the recent detection of GW170608 which hosts a sBH of $7^{+2}_{-2}\,\mathrm{M}_\odot$, the mass function of coalescing sBHs extends, bracketing uncertainties, between about $5\,\mathrm{M}_\odot$ and about $40\,\mathrm{M}_\odot$[Abbott *et al.* (2017b,d,c)].

Theoretical studies (Heger and Woosley, 2002) show that the mass of a sBH, $M_\bullet^{\rm sBH}$, depends not only on the mass of the progenitor star on the zero-age main sequence $m_{\rm zams}$, but also on the metallicity $Z$, and degree of (differential) rotation. The metallicity controls the extent of mass loss by winds during stellar evolution (Vink, 2008). The sBH may form through

*Through this chapter, BHs are described by the Kerr family of solutions and carry *no hair*. The modulus of the spin vector $\mathbf{S}$ is expressed in terms of the dimensionless spin parameter $\chi_{\rm spin} \equiv |\mathbf{S}|c/(GM^2)$, where $M$ is the BH mass. According to the cosmic censorship conjecture $|\chi_{\rm spin}| \leq 1$. If this condition is violated, a naked singularity appears and General Relativity would not be able to describe the exterior solution nor the dynamics of BHs in Nature.

| Abbreviation | full name |
|---|---|
| BH | black hole |
| QSO | quasi-stellar object |
| EM | electromagnetic |
| GW | gravitational waves |
| sBH | stellar-origin black hole |
| SMBH | supermassive black hole |
| IMBH | intermediate-mass black hole |
| MBH | massive black hole |
| PPI | pulsational pair-instability |
| PISN | pair instability supernova |
| AGN | active galactic nuclei |
| MDO | massive dark object |
| GR | General Relativity |
| ISCO | innermost stable orbit |
| BBH | binary black hole |
| PN | post-Newtonian |
| NR | Numerical Relativity |
| SNR | signal-to-noise ratio |
| IMRI | intermediate mass ratio inspirals |
| EMRI | extreme mass ratio inspiral |
| IMF | initial mass function |

the fall back of bound material onto a hot-neutron star driven unstable above its maximum mass during the supernova explosion, or through direct collapse. Consensus is rising that heavy sBHs form in sub-solar metallicity environments, and that in general heavier, lower metallicity stars release heavier sBHs (Belczynski *et al.*, 2010; Spera *et al.*, 2015; Belczynski *et al.*, 2016b). However, at increasingly larger stellar masses, pair instabilities and pair instability supernovae create an *edge* in the sBH mass distribution around $40-60\,\mathrm{M}_\odot$ (Giacobbo *et al.*, 2018), and possibly a *gap* between about $40-60\,\mathrm{M}_\odot$ and $120\,\mathrm{M}_\odot$.[†] For stars with $m_{\rm zams} > 220\,\mathrm{M}_\odot$ and metallicities below $0.07\,Z_{\odot}$, the sBH mass function resumes and sBHs with masses in excess of about $120\,\mathrm{M}_{\odot}$ can form, if very massive stars form in Nature. The mass spectrum of currently observed sBHs is sketched in Figure 1.

[†]In metal-poor stars with $Z < 0.7\,\mathrm{Z}_\odot$ (with $\mathrm{Z}_\odot = 0.01524$) and mass $70 < m_{\rm zams}/\,\mathrm{M}_\odot < 150$, pulsational pair-instabilities (PPIs) cause substantial mass loss during explosive oxygen/silicon burning, limiting the maximum mass of the sBH to about $M_\bullet^{\rm sBH} \simeq 40$-$60\,\mathrm{M}_\odot$(Woosley, 2017; Belczynski *et al.*, 2016a; Spera and Mapelli, 2017). Pair instability supernovae (PISNae) appear at lower metallicities, $Z \leq 8\times10^{-3} \simeq 0.5\,\mathrm{Z}_\odot$. In the mass range about $150 < m_{\rm zams}/\,\mathrm{M}_\odot < 220\,\mathrm{M}_\odot$, PISNae lead to the complete disruption of the star, leaving no remnant.

Supermassive black holes (SMBHs) are observed as luminous QSOs and active galactic nuclei (AGN), fed by accretion of gas (Merloni, 2016), or as massive dark objects (MDOs) at the centre of quiescent galaxies by tracing the stellar and/or gas dynamics in the nuclear regions (Kormendy and Ho, 2013). The SMBH mass spectrum currently observed extend from about $5\times10^4\,\mathrm{M}_{\odot}$ (the SMBH in the galaxy RGG118 (Baldassare *et al.*, 2015)) up to about $1.2\times10^{10}\,\mathrm{M}_{\odot}$ (SDSS J0100+2802 (Wu *et al.*, 2015)), as illustrated in Figure 1.

Observations suggest that SMBHs have grown in mass through repeated episodes of gas accretion and (to a minor extent) through mergers with other BHs. This complex process initiates with the formation of a *seed* BH of yet unknown origin. The concept of seed has emerged to explain the appearance of a large number of SMBHs of billion suns at $z\sim 6$, shining when the universe was only 1 Gyr old (Jiang *et al.*, 2016). In addition, the discovery of the correlation (Kormendy and Ho, 2013) between the SMBH mass $M_\bullet$ and the stellar velocity dispersion $\sigma$ in nearby spheroids, extended to low values of the velocity dispersion, predicts BH masses of $M_\bullet\sim 10^{4.6}\,\mathrm{M}_{\odot}$, at $\sigma$ as low as $40\,\mathrm{km\ s}^{-1}$ (typical of nuclear star clusters) (van den Bosch, 2016). Furthermore, the comparison between the local SMBHs mass density, as inferred from the $M_\bullet-\sigma$ relation, with limits imposed by the cosmic X-ray background light, resulting from unresolved AGN powered by SMBHs in the mass interval between $10^{8-9}\,\mathrm{M}_{\odot}$, indicates that radiative efficient accretion played a large part in the building of SMBHs below $z\sim 3$, and that information is lost upon their initial mass spectrum (Marconi *et al.*, 2004). Thus, SMBHs are believed to emerge from a population of seeds of yet unconstrained initial mass, in the range from about $100\,\mathrm{M}_{\odot}$ to about $10^{4-5}\,\mathrm{M}_{\odot}$. BHs in this mass interval are customarily referred to as *intermediate mass black holes* (IMBHs).[‡]

The word *seed* embodies two concepts. Seeds are IMBHs that form "early" in cosmic history (at redshift $z\sim 20$, when the universe is only 180 Myr old). They form in metal-poor environments under rare conditions, and grow over cosmic time by accretion and mergers. Some of these seeds are able to grow to giant sizes, of up to $10^8-10^{10}\,\mathrm{M}_\odot$, enabling the presence of luminous QSOs during the epoch of cosmic reionization.

[‡]Here on, we refer to as *seed BHs* those IMBHs that grow to give origin to the population of SMBHs. With *BH seeds* (or better BH progenitors) we refer instead to those stellar configurations, which through core collapse, evolve into IMBHs (or seed BHs) which terminate their evolution as SMBH.

Different formation channels have been proposed for the seeds (Volonteri, 2010; Schleicher *et al.*, 2013; Latif and Ferrara, 2016). *Light seeds* refer to IMBHs of about $100\,\mathrm{M}_\odot$ and up to $\sim 10^3\,\mathrm{M}_\odot$. Seeds of about $100\,\mathrm{M}_\odot$ (above the instability gap) are the relics of the first generation of massive stars of zero or very low-metallicity (see chapter **??**), while seeds of $\sim 10^3\,\mathrm{M}_\odot$ likely arise from runaway collisions of massive stars in dense star clusters of low metallicity (Mapelli, 2016; Devecchi *et al.*, 2012, see chapter **??**), or from repeated mergers of sBHs in star clusters subjected to gas-driven evolution (Lupi *et al.*, 2014).

*Heavy seeds* refer instead to IMBHs of about $10^{4-5}\,\mathrm{M}_\odot$ resulting from the monolithic collapse of massive gas clouds, forming in $T_{\rm vir}\gtrsim 10^4$ K metal-free halos exposed to an intense $\mathrm{H}_2$ photo-dissociating ultraviolet flux (Latif *et al.*, 2013; Dijkstra *et al.*, 2014; Habouzit *et al.*, 2016; Regan *et al.*, 2017, see chapter **??**).[§] The seeds of the first SMBHs are still elusive to the most instruments that exist today, preventing us to set constraints on their nature.

In the current cosmological framework, and according to recent studies (Valiante *et al.*, 2018) that we take here to set a logical flow, light and heavy seeds form around $15 < z < 18$, inside halos of comparable masses of about $10^8\,\mathrm{M}_\odot$, and evolve in isolation for their first 150-200 Myrs. Over this lapse time, the halos are enriched of baryons through gas inflows from the external medium and seeds grow via Bondi accretion (with accretion rates $\dot{M} \propto M_\bullet^2$) or at (super-) Eddington rates ($\dot{M} \propto M_\bullet$). Then halo-halo multiple mergers begin at $z \sim 10$. Seed BHs are necessarily a *transient* population of objects and inferring their initial mass function from observations is possible only as long as they preserve some features acquired at birth. As an illustration of this concept, Figure 2 shows the mass distribution of both light and heavy seeds before halos experience multiple mergers (Valiante *et al.*, 2018). Isolated heavy seeds are accreting gas more efficiently than their lighter counterparts. Both populations grow, but as heavy seeds grow faster, then the two population remain distinct over time, and retain information of their origin. As a result of efficient growth and radiative feedback, the host galaxies of heavy seeds experience a less intense star formation activity, and have lower stellar content and lower metallicities. Below $z < 10$, the fraction of isolated systems dramatically decreases to less than 2% (20%), for heavy (light) seed hosts, as both major and minor mergers occur among halos. Thus, below this redshift *binary seed BHs* might form through halo-halo mergers. As seeds grow via mergers and accretion, it is extremely difficult to infer their initial mass spectrum, as they evolve into massive BHs (MBHs) of $10^6 - 10^7\,\mathrm{M}_\odot$, and some to SMBHs ($10^8 - 10^{10}\,\mathrm{M}_\odot$).

[§]Heavy seeds might also form in major gas-rich galaxy mergers over a wider range of redshifts, as mergers trigger massive nuclear inflows (Mayer *et al.*, 2015).

To unveil seeds and track their early growth we need to have access to an overwhelmingly wide cosmological volume and have the capability of measuring at the earliest possible redshifts, their masses and spins. How can this be possible?

According to General Relativity (GR), *BHs of any flavor captured in binaries are loud sources of GWs at the time of their merging. Their GW signal contains direct information on the individual masses and spins of the BHs, and on the source luminosity distance.* Thus, detecting binary BHs through their GW signal over a wide range of masses offers a unique way to learn about how and when seeds form and pair in binaries.

The *Gravitational Wave Universe* is the universe we can sense using GWs as messengers (Colpi and Sesana, 2017; Amaro-Seoane *et al.*, 2017). By observing this universe over a wide interval of GW frequencies, from mHz to kHz, will enable us answering a number of key questions:

- Are sBHs the only elementary building blocks of SMBHs, or is there a "genetic" divide between sBHs and SMBHs, with heavy seeds being the only path to SMBH formation?
- Are there yet undiscovered channels of SMBH formation, calling for the existence of unforeseen, unstable MDOs?
- Can we predict from cosmological models and numerical simulations the mass, spin and mass ratio distributions of seed BHs?
- How fast do seeds grow hand-in-hand with the growth of cosmic structures?
- Can we distinguish at statistical level from the GW data the nature of the seeds, whether light or heavy, and infer if one population dominates over the other?
- How fast information on the two seed populations has been corrupted by the cosmic growth of structures?
- Do seed BHs merge when halos merge?
- How do seed BHs pair in pristine dark matter halos? Is their pairing driven *in-situ* from local dynamics, or induced by halo-halo mergers only?
- What do low redshift MBHs at the centre of dwarf galaxies or IMBHs in dense star clusters tell us about seeds?

This chapter is organized as follows: In § 2 we briefly describe how gas accretion shapes the mass and spin of BHs. § 3 describes shortly the nature of GWs, while § 4 gives an overview of the dynamics of BH binaries and the GW signal they emit. § 5 describes how seed BHs can be detected as GW sources by LISA, and IMBHs by the LIGO-Virgo Observatories in their advanced configurations. In § 6 we show LISA unique capabilities of detecting BHs in the universe over a yet uncharted interval of masses and redshifts, while in § 7 we describe shortly the formation channels of binary BHs. Coalescence rates for heavy seeds are given in § 8. Before concluding, we explore briefly in § 9 the possibility of discovering IMBHs of $10^2-10^3\,\mathrm{M}_\odot$ retained in globular clusters or/and in nuclear star clusters, and MBHs between $10^4-10^6\,\mathrm{M}_\odot$ in dwarf galaxies by detecting their multiband GW signal.

## 2. Accreting black holes

Seed BHs are *transitional* objects as they are expected to undergo episodes of accretion and mergers that change their mass and spin. Accretion may occur via an optically thick, geometrically thin disc which carries mass and angular momentum, transferred to the BH at the innermost stable circular orbit (ISCO) close to the BH event horizon. The BH spin thus changes in norm and direction (Bardeen and Petterson, 1975). Phases of super-Eddington accretion from a thick disc are possible, which bring to fast mass growth of the seeds in gas-rich gravitational potential wells (Lupi *et al.*, 2016). Accretion at a luminosity $L$ equal to a fraction $f_{\rm E}$ of the Eddington luminosity $L_{\rm E}$ enhances the BH mass on an $e-$folding time $\tau_{\rm BH,acc}=4.7\times10^8\eta/[f_{\rm E}(1-\eta)]$ yr, where $\eta$ is the accretion efficiency, which varies from 0.057 (for $\chi_{\rm spin}=0$) to 0.151 (for $\chi_{\rm spin}=0.9$) and 0.43 ($\chi_{\rm spin}=1$, for a maximally rotating BH). Mass accretion and spin are tightly coupled. A non-rotating (maximally-rotating) BH is spun-up (spun-down) to $\chi_{\rm spin}=1$ (to spin zero) after increasing its mass by only a factor $\sqrt{6}$ ($\sqrt{3/2}$) for prograde (retrograde) accretion. If accretion comes in clumps of matter randomly oriented, after a sufficiently long time the BH spin is expected to be low (as retrograde accretion carries a higher angular momentum then prograde accretion). Instead, episodes of coherent accretion in a stable direction leave behind a highly spinning BH (King *et al.*, 2005; Perego *et al.*, 2009). Measuring the spin of coalescing seeds, IMBHs and MBHs in close binaries bring precious information on the way

BHs have accreted gas in the host galaxy (Sesana *et al.*, 2014).

## 3. Gravitational waves in a nut-shell

According to GR, a GW is a perturbation of the metric tensor $h_{\mu\nu}$ in the 4-dimensional spacetime, which propagates at the speed of light over a stationary (flat) background. In the traceless and transversal gauge, the perturbation $h_{\mu\nu}$, orthogonal to the propagation direction, has vanishing space-time and time-time components so that the GW is characterized by only two linearly independent propagating degrees of freedom, called the cross and plus polarizations, and denoted as $h_\times$ and $h_+$ respectively. A single detector can measure a linear combination, called *strain amplitude*, $h(t) = F_+ h_+(t) + F_\times h_\times(t)$, where $F_+$ and $F_\times$ are the antenna patterns of the detector, which depend on the source sky position (Sathyaprakash and Schutz, 2009).

Gravitational radiation is sourced by non-symmetrically accelerated masses (the "charges" of gravity), and tracks the large scale coherent motions of cosmic bodies. Thus, there is a similarity with the process of emission of EM radiation from accelerated elementary charged particles. But, no gravitational dipole radiation is emitted, as the time derivative of the mass dipole moment $\mathbf{D} = \sum m_i \mathbf{r}_i$ of a closed system is the total linear momentum which is conserved, i.e. $\ddot{\mathbf{D}} = 0$. Hence, at the leading order in the multipole expansion, gravitational radiation is *quadrupolar*, where the quadrupole moment is of order $Q \sim \varepsilon M R^2$ for a source of mass $M$, size $R$, with $\varepsilon$ describing the extent of the non-symmetric deformation. Einstein's linearized field equations establish a relation between $Q$ and the strain amplitude $h \propto \ddot{Q}$: a back of the envelope calculation gives $h \sim (R/d)(R_{\rm G}/R)(v/c)^2$, where $R_{\rm G} = 2GM/c^2$ is the Schwarzschild radius associated to the mass $M$, $d$ the source distance, and $v$ the velocity of the asymmetric mass motion ($G$ and $c$ denote the gravitational constant and the speed of light). A compact self-gravitating source ($R \gtrsim R_{\rm G}$) moving at relativistic speed ($v \lesssim c$) has $h \sim GM/(c^2 d)$. Binary black holes (BBHs) at the time of their coalescence have separations $R$ comparable to about $2R_{\rm G}$ (with $M_{\rm B}$ the total mass of the binary) and thus $h \sim 2GM_{\rm B}/(c^2 d)$. In typical astrophysical conditions, $h \sim 10^{-22} - 10^{-19}$. The frequency of the GW is related to the inverse of the dynamical time associated to the non-symmetric motion, $f_{\rm gw} \sim \sqrt{R_{\rm G}/R}\,(c/R)$. In the case of a BBH coalescence, $f_{\rm gw}$ is close to $c^3/(GM_{\rm B})$, as described in section 4.

## 4. Linking the black hole binary dynamics to the GW signal

In GR, a binary system comprising two masses emits GWs and as GWs carry away orbital energy and angular momentum, the binary is fated to coalesce. Calculating the dynamics and radiative properties of two coalescing BHs requires solving Einstein's non-linear field equations in the framework of the so called "geometro-dynamics" where spacetime is foliated into a series of surfaces that allow the notion of time (Lehner and Pretorius, 2014).

The coalescence of a BH binary in vacuum has the beauty of being one of the elementary processes in Nature. It is a well posed problem characterized by a relatively small set of intrinsic parameters: the mass ratio $q \equiv m_2/m_1$, where $m_1$ and $m_2$ are the masses of the two BHs in the binary, and the BH individual spins $\mathbf{S}_1$ (equal to $\chi^1_{\rm spin} G m_1^2/c$) and $\mathbf{S}_2$ defined at large (initial) separations. The initial semi-major axis of the relative orbit (or total energy $E_{\rm B}$), and the modulus of the orbital angular momentum $\mathbf{L}$ (whose direction can be taken as reference axis due to the isotropy of space) define the initial condition and the nature of the orbit (whether it is a head-on collision or a gentle inspiral from an initially bound orbit). The merger product is a new Kerr black hole of mass $M_{\rm f}$ and spin $\mathbf{S}_{\rm f}$. There is no intrinsic scale in vacuum Einstein's gravity. Thus any solution can be re-scaled knowing the binary mass $M_{\rm B} = m_1 + m_2$ under study.

Coalescing BHs produce a universal GW signal which contains unique information on the BH individual masses and spins that source the gravitational field and its vibrations. The GW signal is as a progression of 3+1 phases:

• The *late-inspiral* refers to the phase in which the two BHs can still be considered as structureless and their dynamics and emission can be described by Post Newtonian (PN) theory (Blanchet, 2014). The two BHs in this phase have small orbital velocities compared to the speed of light $c$. This phase, which is the longer lasting, is crucial in obtaining first estimates of the binary system's parameters. If the BHs carry high spins, spin-orbit and spin-spin couplings lead to in-plane and off-plane spin precessions around the total angular momentum (Schnittman, 2004), and the GW encodes these changes (Santamaría *et al.*, 2010).

• The *merger* refers to the non linear plunge-in phase, coalescence and early post-merger. Moving at around one third of the speed of light, the two highly deformed BH spacetimes are described in the realm of Numerical Relativity (NR) as the faithfulness of the PN expansion decreases. The plunge refers to the rapid increase in the magnitude of the inward radial velocity leading to merger, and initiates when the two BHs attain a distance comparable to ISCO, defined here as $R_{\rm ISCO} = 6GM_{\rm B}/c^2$. NR simulations are successful in tracing the non-linear dynamics of the highly distorted vacuum spacetime (Lehner and Pretorius, 2014).

• The *ring-down* is the phase describing the rapid settling of the new BH into a quiescent, stationary state described by the Kerr family which is attained after the GWs carry away any residual spacetime deformation. The radiation consists of a superposition of quasi-normal modes whose frequencies and decaying timescales depend only on the mass and spin of the BH, according to the "no-hair" theorem (Berti *et al.*, 2016). This phase is tracked using NR simulations and/or perturbation theory.

• A fourth phase of *recoil* exists, as the new BH that forms experience a natal kick (Zlochower *et al.*, 2011). The gravitational recoil emerges when the two BHs are not symmetric, as in this case GWs carry away linear moment. The asymmetry can be due to unequal masses, unequal spins, or a combination of the two. For non spinning BHs, the maximum recoil is of $\sim 175\,{\rm km\ s}^{-1}$ when the mass ratio is $q \sim 0.195$. But generic binaries with in-plane spin components may lead to much higher recoil velocities, up to $\sim 4000\,{\rm km\ s}^{-1}$ which arise when equal-mass, maximally spinning BHs merge with spins in the orbital plane equal in magnitude and opposite in direction.

In the next section we focus on key properties of the GW signal in the inspiral phase, particularly relevant when searching for high redshift seed BHs.

### 4.1. *The inspiral and chirp of black holes in binaries*

The GW signal during the inspiral phase can be represented as a sequence of quasi-closed orbits where the semi-major axis $a$ and eccentricity $e$ both decrease with time, as in GR, orbital energy is carried away by GW radiation faster than orbital angular momentum. Little eccentricity is indeed left at the time of coalescence, unless binaries form in triple interaction (Bonetti *et al.*, 2017). For this reason and for quest of simplicity, we focus only on *circular* binaries. Under this assumption, the GW signal is, to leading order, a monochromatic wave of frequency $f_{\rm gw} = 2f_{\rm K} = \pi^{-1}(GM_{\rm B}/a^3)^{1/2}$ in the source rest frame. $f_{\rm gw}$ depends on the square root of total mass of the binary $M_{\rm B}$ and on $a^{-3/2}$. Thus, depending on the separation $a$, BBHs can emit at arbitrarily low frequencies, and only when approaching coales-cence, i.e. when $a$ approaches $R_{\rm ISCO}$, the frequency of the GW attains its maximum.

The orbital energy of a binary is a function of $a$, and can be expressed in terms of $f_{\rm gw}$: $E_{\rm B} = -\nu G M_{\rm B}^2/2a = -[\pi^{2/3}/2]\nu M_{\rm B}(GM_{\rm B})^{2/3}(f_{\rm gw}^{2/3})$, where $\nu$ is the *symmetric mass ratio*, $\nu = m_1 m_2/M_{\rm B}^2 = q/(1+q)^2$ (equal to 1/4 for an equal mass binary and invariant under the change of the mass ratio $q = m_2/m_1$ in $q^{-1}$). As GWs carry away energy, $\dot{E}_{\rm B} < 0$, the binary contracts. During inspiral, the semi-major axis decreases, the Keplerian frequency increases and thus $f_{\rm gw}$. To leading order, the key (orbit-averaged) equation that rules the binary contraction during the quasi-adiabatic, long-lived inspiral, is

$$\dot{E}_{\rm B} = -P_{\rm GW} \equiv -\frac{32}{5}\frac{c^5}{G}\left(\frac{GM_{\rm c}}{c^3}\pi f_{\rm gw}\right)^{10/3} \tag{1}$$

where the right-hand side gives the radiated power in GWs, $P_{\rm GW}$, for a circular binary.[¶] Equation (1) contains the so called binary *chirp* mass

$$M_{\rm c} \equiv \frac{(m_1 m_2)^{3/5}}{(m_1+m_2)^{1/5}} = \nu^{3/5} M_{\rm B} = \frac{q^{3/5}}{(1+q)^{6/5}} M_{\rm B}^{2/5}, \tag{2}$$

i.e. a combination of the individual BH masses, expressed in terms of the symmetric mass ratio $\nu$.

As $E_{\rm B}$ and $f_{\rm gw}$ can be expressed in terms of the semi-major axis $a$, equation (1) can be recast to give the rate of change of $\dot{a}$. It is then a simple exercise to determine the *coalescence time*, defined as the time it takes a BH binary, driven by GW losses, to reach null separation (in the point mass approximation), knowing that the binary formed with an initial semi-major axis $a_\circ$, set by "astrophysical" conditions. For a circular binary

$$t_{\rm coal} = \frac{5}{256}\frac{c^5}{G^3}\frac{{a_\circ}^4}{\nu\, M_{\rm B}^3} = \frac{5\cdot 2^4}{256}\frac{1}{\nu}\frac{GM_{\rm B}}{c^3}\tilde{a}_\circ^4, \tag{3}$$

where $\tilde{a}_\circ$ is the binary semi-major axis in units of $R_{\rm G} = 2GM_{\rm B}/c^2$. [If the binary is eccentric with initial eccentricity $e_\circ$, the coalescence time is shorter by a factor $G(e_\circ)(1-e_\circ^2)^{7/2}$, with $G(e_\circ)$ varying from 1 (for $e_\circ = 0$) to 1.80 (for $e_\circ = 1$).]

Due to the weakness of the gravitational coupling constant and the steep dependence of $\tau_{\rm coal}$ on $a_\circ$, the coalescence time can be very long, unless the binary is initially extremely eccentric. A (circular) binary of $M_{\rm B} = 10^5\,{\rm M}_{\odot}$ ($10^3\,{\rm M}_{\odot}$) reaches coalescence in 0.27 Gyrs, corresponding to the cosmic time at redshift $z \sim 15$, if the two BHs have a separation of $a_{\circ} \sim \nu^{1/4}\,1.5\times 10^4 R_{\rm G}$ ($\nu^{1/4} 4.84\times 10^4 R_{\rm G}$), corresponding to $\nu^{1/4}\,30$ AU ($\nu^{1/4}\,0.1$ AU). Reaching these separations requires the presence of non-GW dissipative processes, active when the binary interacts with the environment or when the progenitor stars undergo a fine-tuned evolution, as discussed in section 7.

[¶]We defer the reader to the book by M. Maggiore, *Gravitational Waves* (Oxford University Press), for a detailed derivation of all the equations presented in this Chapter.

Coalescing seed BHs are at cosmological distances. Thus the GW signal is affected by the expansion of the universe. If $z$ is the cosmological redshift of the source of GWs, it is possible to show that (i) the observed frequency, denoted for simplicity as $f$ here-on, is redshifted with respect to the frequency $f_{\rm gw}$ as measured in the source frame, $f = f_{\rm gw}/(1+z)$; (ii) the chirp mass in the source frame $M_{\rm c}$ is replaced by the redshifted chirp mass $\mathcal{M}_{\rm c} = (1+z)M_{\rm c}$; (iii) the source distance $d$ is identified with the luminosity distance $d_L$; and (iv) the time to coalescence as measured in the observer frame is $\tau_{\rm coal} = (1+z)t_{\rm coal}$. Note that the combination $\mathcal{M}_{\rm c}f = M_{\rm c}f_{\rm gw}$ is redshift independent.

Equation (1) can be recast to determine the rate of change of the frequency of the GW, $f_{\rm gw}$, during the inspiral phase, in the source frame. If one includes PN terms and refers to the rate of change of $f$ in the observer frame, the frequency is chirping at a rate

$$\dot{f} = \frac{96}{5}\pi^{3/8}\left(\frac{G\mathcal{M}_{\rm c}}{c^3}\right)^{5/3} f^{11/3}\left[1+\mathcal{D}^{\rm PN}\right]. \tag{4}$$

where $\mathcal{D}^{\rm PN}$ is the PN correction, up to the desired order higher than leading (Blanchet, 2014). $\mathcal{D}^{\rm PN}$ depends on the initial phase and coalescence time. It can be expanded analytically in powers of $\nu$ and of $\mathcal{M}_{\rm c}f$. Spin effects can also be inserted in $\mathcal{D}^{\rm PN}$ [Blanchet (2014)].

The frequency evolution of the GW is determined by the chirp mass $\mathcal{M}_{\rm c}$, according to equation (4). Thus, measuring $\dot{f}$ provides directly the measure of the redshifted chirp mass associated to the binary. The PN corrections present in $\mathcal{D}^{\rm PN}$ help lifting the degeneracy in the intrinsic parameters of the source present in the signal, and in particular the individual masses in the observer frame.

If we neglect for simplicity the PN term in equation (4), we can estimate the residence time of an inspiraling BBH measured by the observer's clock, when it enters the sensitivity window at a frequency $f_{\rm in}$ and exits at $f_{\rm out}$:

$$\tau_{\rm det} = \frac{5}{256\,\pi^{8/3}}\frac{1}{\nu}\left(\frac{c^3}{G\mathcal{M}_{\rm c}}\right)^{5/3}\left(\frac{1}{f_{\rm in}^{8/3}} - \frac{1}{f_{\rm out}^{8/3}}\right). \tag{5}$$

Equation (5) shows that an unequal mass binary with $\nu \ll 1$, and/or a binary with a smaller chirp mass stay in band much longer covering many cycles. A back of the envelope calculation shows that the number of cycles that a BBH covers before merging scales as $\mathcal{N}_{\rm cycles} \propto \nu^{-1} M_{\rm B}^{-5/3} f_{\rm in}^{-5/3}$.

The chirp signal contains further information on the binary, as it is accompanied by an increase of the amplitude $h$ of the GW. According to GR, the GW signal from an inspiraling binary can be decomposed in its plus and cross polarization states, which take the following form in the Fourier domain:

$$\tilde{h}_{+}(f) = \left(\frac{5}{12}\right)^{1/2}\frac{1}{\pi^{2/3}}\frac{c}{d_L}\left(\frac{G\mathcal{M}_{\rm c}}{c^3}\right)^{5/6} e^{i\Phi_{+}^{\rm PN}(f)}\frac{1}{f^{7/6}}\left(\frac{1+\cos^2\iota}{2}\right) \tag{6}$$

$$\tilde{h}_{\times}(f) = \left(\frac{5}{12}\right)^{1/2}\frac{1}{\pi^{2/3}}\frac{c}{d_L}\left(\frac{G\mathcal{M}_{\rm c}}{c^3}\right)^{5/6} e^{i\Phi_{\times}^{\rm PN}(f)}\frac{1}{f^{7/6}}\cos\iota \tag{7}$$

where $\Phi_{+}^{\rm PN}(f) = \Phi_{\times}^{\rm PN}(f) - \pi/2$ are the phases, computed at the desired PN order, and $\iota$ is the binary inclination angle, i.e. the angle between the orbital angular momentum $\mathbf{L}$ and the line of sight. The polarization of the GW has a direct relationship to the motions of the BHs projected on the observers sky plane, and the $h_{+}/h_{\times}$ gives directly the inclination angle $\iota$. When the binary is seen edge-on ($\iota = \pi/2$), $h_{\times} = 0$, radiation has $+$ polarization only, since from the observer's view the motion of the two BHs projected on the sky is linear. When $\iota = 0$, the binary is seen face-on and the BHs execute a circular motion in the sky. In this case both polarization components have equal amplitude and are out of phase by $\pi/2$.

If the GW detector (or a network of detectors) enables the measure of the two independent polarization states of a GW, from the chirp signal (eq. 4, 6, and 7) one can infer the luminosity distance $d_L$ of the source. From the ratio $h_{+}/h_{\times}$ we obtain the inclination angle $\cos\iota$, from the frequency chirp the redshifted chirp mass, so that from the measured value of $h_{+}$ (or $h_{\times}$) we can read off $d_L$. Coalescing BBHs are gravitational standard candles, or as often said, *standard sirens.* The application of the luminosity distance-redshift relation ($d_L$ versus $z$) from the current cosmological model provides the redshift of the source $z_{\rm source}$ and the inference of the source-frame physical parameters. But, if an EM counterpart is associated to and identified with the GW source, it is then possible to have a direct measure of $z_{\rm source}$ and this allows for an independent measure of the Hubble expansion parameter to a certain degree of accuracy (Tamanini *et al.*, 2016).

### 4.2. *The merger and the newly formed black hole*

The *merger* occurs in the strong field regime when the spacetime of the two plunging BHs is violently changing, and NR simulations are the only tool to track the dynamics and extract the GW signal in the far field. The observation of the first coalescing BBHs (Abbott *et al.*, 2016a) has shown that the GW signal is extremely simple at merger: after the chirp, the amplitude of the GW is observed to decay rapidly, as predicted by NR simulations and perturbation theory. The frequency derivative is no longer described by equation (4), as it coasts to a finite value, when the new BH forms. This maximum frequency, derived from NR simulations, is close to twice the Keplerian frequency at the ISCO for a binary of total mass $M_{\rm B}$,

$$f_{\rm gw,max} \sim \frac{c^3}{\pi 6^{3/2} G M_{\rm B}} = 4.4 \times 10^3 \left(\frac{{\rm M}_\odot}{M_{\rm B}}\right) {\rm Hz}. \tag{8}$$

The simple scaling of $f_{\rm gw,max}$ with $M_{\rm B}$ shows that detecting coalescing sBHs of $10\,{\rm M}_\odot$ requires interferometers operating at frequencies $10^4$ times higher than those requested to detect IMBHs in a binary with $M_{\rm B} = 10^5\,{\rm M}_\odot$.

The merger of BBHs carries tremendous luminosity of the order of $L_{\rm GW} \approx \nu\ (c^5/G) \sim 3.6\times\ 10^{59}\ \nu$ erg s$^{-1}$ which is independent on $M_{\rm B}$, as in GR mass $M_{\rm B}$ is a length, and time has same dimension. This huge luminosity is emitted for a short time lapse, in the immediate vicinity of the merger proper. What depends on the mass scale is the total energy radiated away in GWs. It arises from two sources: the gravitational binding energy liberated during the inspiral and plunge, and the energy present in the geometry of the deformed, rotating relic before it settles into a new Kerr BH. As rule of thumb, the radiated energy is approximately equal to the binding energy at ISCO of a particle with reduced mass $\mu$ (where $\mu = \nu M_{\rm B}$), and can account for $3.8\% - 42\%$ of $\mu c^2$ depending on the spin of the primary (i.e. heavier) BH and whether the orbit is prograde or retrograde, relative to the spin direction.

If the inspiral, merger and ringdown of a coalescing binary is fully in band and the source is detected in both polarization states with a high signal-to-noise-ratio, then it is possible to carry on highly accurate measurements of the intrinsic and extrinsic parameters of the source, as the luminosity distance, the individual masses and spins of the BHs, and the final mass and spin of the new BH, in the source frame.[‖] Fitting formulae instructed by NR simulations exist (Barausse *et al.*, 2012) that help inferring the final mass of the new BH, given approximately by the difference between the initial mass $M_{\rm B}$ and the energy carried away by the GWs. Likewise, the final spin $\mathbf{S}_{\rm f}$ can be viewed to lowest order as the sum of the orbital angular momentum $\mathbf{L}$ at ISCO plus the intrinsic angular momenta of the two BHs. For equal-mass non spinning BBHs, the final spin is $\chi_{\rm spin,f} = 0.6865$.

### 4.3. *Extracting the signal from the noise*

Match filtering is the technique used to extract the source signal from the noise. The signal-to-noise ratio (SNR) (or $\rho$) of a source is computed as

$$({\rm SNR})^2 = \int_0^\infty d\ln f \, \frac{|2\tilde{h}(f)\sqrt{f}|^2}{S_{\rm noise}} \tag{9}$$

where $S^{1/2}_{\rm noise}$ is the *spectral strain sensitivity* or *spectral amplitude* (with dimension $\rm Hz^{-1/2}$) describing the noise in the detector, and $\tilde{h}(f) = F_+\tilde{h}_+(f) + F_\times\tilde{h}_\times(f)$, (in units of $\rm Hz^{-1}$). The SNR accumulates over the observing time, until the source rapidly fades away, as in the case of a coalescing BH binary, or exits from the observing bandwidth. Customarily, $(fS_{\rm noise})^{1/2}$ and $f\tilde{h}(f)$ are plotted in the same diagram as function of frequency $f$, to describe the sensitivity window of an interferometric experiment and the strength of the source signal, respectively. During BBH inspiral $f\tilde{h}(f) \propto f^{-1/6}$ decreases with increasing $f$, as the source completes less cycles in a given frequency interval. But, the decrease in $f\tilde{h}$ is compensated by the increase in the SNR, which accumulates over time up to $f_{\rm max}$ or $f_{\rm out}$. Typically, merging BBHs cover from a few and up to many cycles $\mathcal{N}_{\rm cycles} \sim O(10) - O(1000)$ depending on their mass and on the value of the frequency $f_{\rm in}$ at the time of entrance in the interferometer.

## 5. High redshift binary seed black holes and low redshift intermediate mass black holes

The *Gravitational Wave Universe* is the universe we can explore using GWs as new messengers over a wide range of frequencies. Coalescing *heavy* seeds

[‖] A current and powerful tool used to carry on parameter estimation is the Effective One Body (EOB) theory, where re-summed PN inspiral waveforms are smoothly attached to quasi-normal ringdown modes via a transition function calibrated by NR (Buonanno and Damour, 1999; Buonanno and Sathyaprakash, 2014).

and MBHs of up to a few $10^7\,\mathrm{M}_\odot$ are the targets of the Laser Interferometer Space Antenna (LISA) operating in the 0.1 mHz-0.1 Hz interval [Amaro-Seoane *et al.* (2013, 2017)]. *Light* seeds are the targets of the third generation of ground-based detectors (III GBI) as the Einstein Telescope [Sathyaprakash *et al.* (2012); Abbott *et al.* (2017a)], sensitive between about 1 Hz (or a few Hz) and several $10^3$ Hz, that will let us detect the signal from sBHs and IMBHs in binaries out to high $z \sim 5-6$. sBH binaries with total masses between $60\,\mathrm{M}_\odot$ and $300\,\mathrm{M}_\odot$ can sweep through the LISA frequency domain years to months before coalescing in the kHz bandwidth [Sesana (2016)]. Advanced LIGO and Virgo interferometers operating in the 10-$10^3$ Hz interval [Abbott *et al.* (2016b)] will have the capability of detecting binary IMBHs of $300\,\mathrm{M}_\odot$ out to $z \sim 1$, setting the conditions for the early search of IMBHs in the universe.

Figure 3 shows the GW signal $f\tilde{h}$ from seed BHs in circular, equal-mass binaries with (source-frame) total mass $M_\mathrm{B} = 10^5\,\mathrm{M}_\odot$ (solid lines) and $10^4\,\mathrm{M}_\odot$ (dashed lines) in blue and light blue colors as they sweep across the LISA band. The signal is shown varying the redshift of the source from $z \sim 13$ down to $z \sim 3$. For BH binaries with $M_\mathrm{B} = 10^5\,\mathrm{M}_\odot$, the coalescence signal is in band enabling accurate source parameter estimations. For BH binaries of $10^4\,\mathrm{M}_\odot$, LISA can only detect the inspiral phase and measure the chirp mass and precession effects if spin are high and misaligned. These seeds are sufficiently light to attain their maximum frequency out of the LISA sensitivity band. Their signal stands above the bucket of the sensitivity curve with accumulated SNR of $\lesssim$50, typically. Figure 3 shows also the signal (solid lines in yellow and dark green colors) from circular, equal-mass binaries with source-frame total mass $M_\mathrm{B} = 300\,\mathrm{M}_\odot$ as they transit across the LISA sensitivity band during the inspiral phase and enter the LIGO-Virgo band where they merge. Only relatively close-by ($z < 1$) IMBHs in equal mass binaries with individual mass of $\sim 150\,\mathrm{M}_\odot$ (just above the pair-instability gap) can be detected both by LISA and LIGO-Virgo. Figure 4 shows the range where both space and Earth-based interferometers can operate jointly in the search of heavy sBHs and IMBHs, in the mass interval between $10\,\mathrm{M}_\odot$ up to several $10^3\,\mathrm{M}_\odot$. At the redshifts $z\lesssim 1$, IMBHs may not represent unevolved light seeds, and may describe a population of IMBHs that form in dense star clusters that build up their present mass, during the cosmic evolution of their host galaxies.

## 6. The black holes of the gravitational wave universe

Figure 5 illustrates the vastness of the cosmic horizon probed by the LISA observatory, in the hunt of heavy seed BHs and of MBHs. Contour lines of constant (sky, inclination and polarization averaged) SNR are shown in the $(M_{\rm B} - z)$ plane, for spinning (non-precessing) BHs in binaries with mass ratio $q = m_2/m_1 = 0.2$. We recall that the SNR scales linearly with $\nu = q/(1+q)^2$, thus binaries of equal mass carry the highest SNR. Astrophysical considerations on BH sinking in galactic mergers suggest that the likely interval for $q$ is between 0.1 and 1 (Colpi, 2014).

As shown in Figure 5, LISA is dominated by high SNR sources, and can reveal BBHs up to $z \sim 20$ in a large mass interval between a few $10^3\,{\rm M}_\odot$ to a few $10^6\,{\rm M}_\odot$, where (mostly heavy) seeds are. Around $z \sim 2{-}3$, during the peak of the star formation rate and AGN activity in the universe (Madau and Dickinson, 2014; Merloni, 2016), LISA can detect the inspiral of IMBHs of about $10^3\,{\rm M}_\odot$, the full signal of BBHs up to $10^7\,{\rm M}_\odot$, and the merger and ringdown signal of eccentric low-redshift BBHs of $10^8\,{\rm M}_\odot$, complementing observations from future EM surveys (Nandra *et al.*, 2013).

BHs cross the LISA bandwidth as mergers are inevitable in our hierarchical universe [Amaro-Seoane *et al.* (2013)]. BBH coalescences pinpoint places where galaxy mergers occur. Thus, LISA provides a new and unique way to explore galaxy clustering. This is illustrated in Figure 5 where we depicted two heavy seeds hosted in their dark matter halos, merging around $z \sim 10$. Both seeds increase their mass before coalescing, fed by gas. We allowed the new BH to grow by accretion to deposit a MBH of a few $10^7\,{\rm M}_\odot$ detectable as AGN during cosmic high noon.

In virtue of the extremely high SNR of most of the events, the BH parameters will be extracted with exquisite precision (Klein *et al.*, 2016). For the loud sources with SNR$>60$, individual redshifted masses can be measured with an error of 0.1% on both components. Even more importantly, the spins of the BHs can be determined to an absolute uncertainty down to 0.01 (Klein *et al.*, 2016), in the loud sources. Information on the astrophysical evolution of massive BHs is encoded in the statistical properties of the observed population. As first illustrated in Sesana *et al.* (2007) observations of multiple BBH coalescences can be combined to learn about their formation path and cosmic evolution. In particular the mass distribution of the ensemble of observed events encodes information on the nature of the first seeds, whereas the spin distribution constrains the primary mode of accretion (whether chaotic or coherent) (Berti and Volonteri, 2008).

The distinctive high SNR of BBH mergers allows for BH spectroscopy, i.e. the direct measure of several frequencies and damping times associated to the quasi-normal modes present in the ringdown signal of the new-born MBH (Berti *et al.*, 2016). This will enable direct precision tests of the "no-hair" theorem in the dynamical sector. Violations of GR prediction indicate new physics, the presence of exotic dark objects and indirect quantum gravity effects (Berti *et al.*, 2015; Cardoso and Pani, 2017).

## 7. Formation channels of binary black holes and time delays

Detecting seed BHs at high redshifts as GW sources poses a stringent requirement: their existence in binaries in a *GW-hard* state, i.e. in binaries with semi-major axis and eccentricity such that the coalescence time $t_{\rm coal}$ given by equation (3) is a fraction of the cosmic time at the redshift of formation, or equivalently a fraction of the then Hubble time. (GW-hard binaries driven by the sole GW emission are the ones described in § 4.) This level of hardness can be computed posing $t_{\rm coal}(a_\circ, e_\circ;\, z) = t_{\rm Hubble}(z)$. The equality provides the values of the semi-major axis $a_\circ$ and eccentricity $e_\circ$ that a binary should have to be GW-hard, at some stage of its evolution.

A binary is in a *GW-soft* state if its coalescence time at formation is longer than the Hubble time. Some never coalesce, other transit to a GW-hard state. The transition from a GW-soft to a GW-hard state is possible and regulated by dissipative processes of non-GW origin. Hydrodynamical or/and stellar torques acting on the binary control this transition and their strength and effectiveness depend on the environment in which the binary is embedded (Colpi, 2014).

In Nature, binaries are expected to form in the GW-soft state, and this leads to the concept of *time delay*, representing the time lapse between the time of formation of the binary $t_{\rm form}$ at a given redshift $z$ and the time of coalescence as measured in the source frame. Long time delays make it possible to detect sBH which formed at a much earlier epoch, when the host, now incorporated in a larger galaxy, was metal-poor. This is the case of GW150914 (Schneider *et al.*, 2017). But detecting seeds at high redshifts requires binaries to form GW-hard.

Little is known about how seed BHs pair to form GW-hard binaries in the high-$z$ universe. They may form in-situ or via halo-halo mergers. Heavy seeds in GW-hard binaries can form in-situ, i.e. within a halo, if the progenitor is a single supermassive star rotating differentially and subjected to the $m = 2$ mode instability. Hydrodynamical simulations, which incorpo-rate the dynamical spacetime evolution, show that an $m = 2$ perturbation nested inside the flow can grow, leading to the formation of a binary of two highly spinning heavy seeds which swiftly merge in a short time (Reisswig *et al.*, 2013). Light seeds forming inside massive ($> 10^7\,\mathrm{M}_\odot$) dense star clusters of low but not-null metallicity (Devecchi *et al.*, 2012), might pair and form GW-hard binaries through the dynamical channel as they gradually harden via transient three-body or four-body encounters, leading to frequent substitutions of the sBH/IMBH companion, large amplitude oscillations of the eccentricity, and finally to coalescence (Mandel *et al.*, 2008; Haster *et al.*, 2016). Intermediate mass ratio inspirals (so called IMRIs) involving a binary system with a sBH and an IMBH may form in Nature and be detectable as GW sources at relatively low $z$. These events however may not represent the mergers of light seed BHs in the high redshift universe, and only deep observations with LISA will let us distinguish among possible pathways.

An alternative avenue of formation of heavy seeds in binaries is via halo-halo mergers whose dynamics is influenced by the evolving cosmological background. At present, cosmological simulation do not have the force resolution to track in full detail the dynamics of the seed BHs embedded in their halos (from $10^3$ parsecs scale down to a few AU), nor their accretion and associated feedback, which are implemented via sub-grid prescriptions at scales of several parsecs. There exists a sub-grid advection scheme in which BHs are artificially dragged toward the halo centre and then forced to merge promptly as the two halos merge. Recent simulations by Tremmel *et al.* (2015), which include fine-tuned dynamics to account for unresolved dynamical friction, find that MBHs from merging dwarf galaxies can spend significant time away from the centre of the remnant galaxy. In future, improving the modeling of seed/IMBH/MBH orbital decay will help in making predictions on the growth, detectability, and merger rates of BHs over a wide mass spectrum and redshifts.

Dedicated, higher resolution numerical simulations of merging galaxies (not in a full cosmological context) indicate that in gas-rich galaxy-galaxy merger the MBHs sink to the centre-most regions of the new galaxy to form a Keplerian binary (when the mass in stars and gas inside the BH orbit becomes comparable to $M_{\mathrm{B}}$), if the two galaxies have comparable mass. In unequal mass mergers, sinking of the BHs is effective if they are surrounded by a massive stellar cusp that formed during the merger. This process enhances the BH effective mass, and stellar dynamical friction becomes more efficient in braking the BH orbits (Callegari *et al.*, 2009). Due to the richness and variety of galaxy encounters, there is the possibility that in gas-rich galactic discs one or both MBHs are scattered away from the disc plane in their dynamical interaction off massive clouds. Alternatively, one MBH can be trapped in a massive cloud, and this leads to a broadening of the sinking times, which vary from a few hundreds of Myrs to a few Gyrs. The BH binary that forms then harden via scattering off individual stars which extract orbital energy and angular momentum (the so called sling-shot mechanism). This process is rapid as long as their is a sufficiently large reservoir of stars on loss cone orbit, in a tri-axial remnant. Processes reminiscent to Type II planet migration (Haiman *et al.*, 2009) can further harden the binary on scales below the mpc if the BHs are surrounded by a circum-binary gas disc. Gravitational gas-dynamical torques are able to taxi the BHs into the GW-driven regime. At present there is only one simulation capable of tracking the full dynamical history of a $10^8\ M_\odot$ BH binary from the kpc scale down to merger proper (Khan *et al.*, 2016). (For a review on the BH dynamics in minor and major galaxy merger we refer to Colpi (2014).)

## 8. Coalescence rates of binary black holes in LISA

The expected rate of BBH coalescences is weakly constrained by current EM observations of galaxy-galaxy mergers with MBHs at their centre. Estimating the rate is extremely difficult as it depends on a variety of processes difficult to model, among which: (i) the mass spectrum of the seeds and their initial occupation fraction in galactic halos; (ii) their BH mass and spin accretion history; (iii) the type of galaxy hosts in which the BHs inhabit; (iv) the dynamics of pairing and hardening of the BHs in the new galaxy halo, depending on the gas and stellar content of the progenitor halos/galaxies and on their orbits in a cosmological context; (iv) the extent of GW recoils which may decrease the occupation fraction of BHs in galaxies that can further merge.

Figure 6 shows the rates expected from three models associated to different seeding mechanisms described in the caption (Klein *et al.*, 2016). The difference in the merger rates among the models is due to the different mass functions and occupation numbers of the light seeds compared to the heavy seeds, and to the strength of the dynamical interactions that the seeds experience with stars and gas in order to pair on the smallest scales where GWs drive the inspiral. The three models bracket the possible range of coalescence rates. Cosmological simulations of the galaxy assem-bly anchored to estimates of the local galaxy-merger-rate predict a few to few-hundred coalescences per year for the LISA events.

## 9. Massive black holes in the near universe and the role of EMRI's LISA sources

Dwarf galaxies are expected to experience relatively quiet merger histories, and thus potentially host the least-massive BHs of the today universe. Extending the local $M_\bullet - \sigma$ correlation and ancillary relations as the $M_\bullet - M_*$ one to low mass galaxies leads to a larger scatter of the data, and a richness in the host galaxy's morphology and kinematics (Kormendy and Ho, 2013). Thus, even in the zero-redshift universe, the knowledge of the low-end of the SMBH mass distribution, and of the occupation fraction in galaxies is incomplete. Often in low mass galaxies, a nuclear star cluster is in place, and a central MBH/IMBH may or may not cohabit the cluster.

A systematic search for the EM signature of an AGN in dwarf galaxies has been carried in Reines *et al.* (2013), who showed interestingly that (even bulge-less) dwarf galaxies host central BHs with median virial masses of $10^5\,\mathrm{M}_\odot$. Whether this mass is close to the seed mass is not predictable. Over the age of the universe, even in the quiet environment of a dwarf one can not exclude a secular growth due to stellar tidal captures and disruption events (Stone *et al.*, 2017). This may lead to the growth of the BH of up to a "minimal" mass $M^{z=0}_{\rm min} \sim 10^5\,\mathrm{M}_\odot$, irrespective of the initial mass, thus erasing information on the seed formation mechanisms (Alexander, 2017). EM observations of nearby galactic nuclei will bring further insight into this process of mass saturation at a minimal level.

LISA will also contribute to increment knowledge on the mass distribution of MBH in nearby galaxies in the range between a few $10^4\,\mathrm{M}_\odot$ up to a few $10^6\,\mathrm{M}_\odot$. Exquisite probes are the so called Extreme Mass Ratio Inspirals (EMRIs) (Gair *et al.*, 2017). An EMRI is a extreme mass ratio binary comprising a MBH and a sBH moving on a highly relativistic, mildly eccentric (0-0.2) orbit. The sBH performs about $10^4 - 10^5$ cycles before crossing the horizon of the large BH, and its orbit displays extreme forms of in plane and off-plane relativistic precession. The waveform encodes information on the masses of the two BHs, the eccentricity of the orbit at plunge and the spin of the large BH to such an extent that it will make it possible to measure deviations from the quadrupole moment of the the central object up to a precision of one part in $10^{-4} - 10^{-3}$ (Gair *et al.*, 2017). This shall enable us to test whether the central object is a MBH or a horizonless object as a boson star, and measure the redshifted mass of the sBH a with a precision of one part in $10^{-5}$. EMRI rates are uncertain and range between 0 and 100 $\mathrm{yr}^{-1}$ (Alexander and Bar-Or, 2017). Loss cone dynamics at the gravitational sphere of influence of the massive MBH suggests that an EMRI is produced out of $\sim 100$ direct plunges, driven by angular momentum changes over the two-body relaxation times. The EMRI's dynamics is extremely sensitive to the stellar distribution around the MBH, so that detecting EMRIs not only will test the nature of the central object but thanks to the accurate mass measurement it can probe the sBH and SMBH mass distribution in Milky Way like galaxies as well as dwarfs, out to a redshift $z \sim 2$. The EMRI signal is just detectable above the bucket of the LISA observatory sensitivity, where there are also the signals from the coalescing seed BHs.

## 10. BH archaeology

Following the first direct detection of GWs by the LIGO Scientific Collaboration and Virgo Collaboration, and the success of the LISA Pathfinder mission, we can firmly claim that we are heading into an era of discoveries with the opening of a new window on the universe: *the GW window*. There is a *fil rouge* connecting sBHs and SMBHs. SMBHs, already in place at redshifts as large as $z \sim 7$, may rise from a unique building block, the BHs of stellar origin, or from many avenues, e.g. from a population of IMBHs still elusive to EM observations. A possibility not envisaged in this chapter is that a fraction of BHs in the mass window between $25 - 100\,\mathrm{M}_\odot$ are of primordial origin (Carr *et al.*, 2017). The discovery of primordial BHs with mass below the maximum mass of a neutron star, would also be ground-braking as it would indicate a formation path of cosmological (non-stellar) origin taking us closer to the Big Bang. LISA shall have the capability of detecting the early-forming heavy seeds and track their growth across all cosmic ages, surveying the low-mass tail of the SMBH mass spectrum. Ground based interferometers in their advanced configurations and the third generation of detectors will push observations of sBHs close to the epoch of cosmic reionization, unveiling the high-mass end of sBHs. With this new field of BH archaeology, we will be able to discover if BHs over a wide interval of masses, from about $10\,\mathrm{M}_\odot$ up to $10^{10}\,\mathrm{M}_\odot$ are the universal outcome of un-halted gravitational collapse, in Nature.

Overall, we can conclude that gravitational waves may provide the perhaps most promising pathway to determine the BH mass function over a large range of redshifts, and it is important to extend current facilities to also probe GWs from black hole mergers in the intermediate or supermassive range. These investigations can and should be complemented with observations of classic EM radiation. The next chapter will thus conclude the book with future prospects from EM observations.

## References

Abbott, B. P., Abbott, R., Abbott, T. D., Abernathy, M. R., Acernese, F., Ackley, K., Adams, C., Adams, T., Addesso, P., Adhikari, R. X. and et al. (2016a). Observation of Gravitational Waves from a Binary Black Hole Merger, *Physical Review Letters* **116**, 6, 061102, doi:10.1103/PhysRevLett.116.061102.

Abbott, B. P., Abbott, R., Abbott, T. D., Abernathy, M. R., Acernese, F., Ackley, K., Adams, C., Adams, T., Addesso, P., Adhikari, R. X. and et al. (2016b). Prospects for Observing and Localizing Gravitational-Wave Transients with Advanced LIGO and Advanced Virgo, *Living Reviews in Relativity* **19**, doi:10.1007/lrr-2016-1.

Abbott, B. P., Abbott, R., Abbott, T. D., Abernathy, M. R., Ackley, K., Adams, C., Addesso, P., Adhikari, R. X., Adya, V. B., Affeldt, C. and et al. (2017a). Exploring the sensitivity of next generation gravitational wave detectors, *Classical and Quantum Gravity* **34**, 4, 044001, doi:10.1088/1361-6382/aa51f4.

Abbott, B. P., Abbott, R., Abbott, T. D., Acernese, F., Ackley, K., Adams, C., Adams, T., Addesso, P., Adhikari, R. X., Adya, V. B. and et al. (2017b). GW170104: Observation of a 50-Solar-Mass Binary Black Hole Coalescence at Redshift 0.2, *Physical Review Letters* **118**, 22, 221101, doi:10.1103/PhysRevLett.118.221101.

Abbott, B. P., Abbott, R., Abbott, T. D., Acernese, F., Ackley, K., Adams, C., Adams, T., Addesso, P., Adhikari, R. X., Adya, V. B. and et al. (2017c). GW170608: Observation of a 19 Solar-mass Binary Black Hole Coalescence, *ApJ* **851**, L35, doi:10.3847/2041-8213/aa9f0c.

Abbott, B. P., Abbott, R., Abbott, T. D., Acernese, F., Ackley, K., Adams, C., Adams, T., Addesso, P., Adhikari, R. X., Adya, V. B. and et al. (2017d). GW170814: A Three-Detector Observation of Gravitational Waves from a Binary Black Hole Coalescence, *Physical Review Letters* **119**, 14, 141101, doi:10.1103/PhysRevLett.119.141101.

Alexander, T. (2017). Stellar Dynamics and Stellar Phenomena Near a Massive Black Hole, *ARA&A* **55**, pp. 17–57, doi:10.1146/annurev-astro-091916-055306.

Alexander, T. and Bar-Or, B. (2017). A universal minimal mass scale for present-day central black holes, *Nature Astronomy* **1**, 0147, doi:10.1038/s41550-017-0147.

Amaro-Seoane, P., Aoudia, S., Babak, S., Binétruy, P., Berti, E., Bohé, A.,

Caprini, C., Colpi, M., Cornish, N. J., Danzmann, K., Dufaux, J.-F., Gair, J. and Hinder, I. et al. (2013). eLISA: Astrophysics and cosmology in the millihertz regime, *GW Notes, Vol. 6, p. 4-110* **6**, pp. 4–110.

Amaro-Seoane, P., Audley, H., Babak, S., Baker, J., Barausse, E., Bender, P., Berti, E., Binetruy, P., Wass, P., Weber, W., Ziemer, J. and Zweifel, P. (2017). Laser Interferometer Space Antenna, *ArXiv e-prints 1702.00786* .

Antonini, F., Barausse, E. and Silk, J. (2015). The Coevolution of Nuclear Star Clusters, Massive Black Holes, and Their Host Galaxies, *ApJ* **812**, 72, doi:10.1088/0004-637X/812/1/72.

Baldassare, V. F., Reines, A. E., Gallo, E. and Greene, J. E. (2015). A 50,000 Solar Mass Black Hole in the Nucleus of RGG 118, *ApJ* **809**, L14, doi: 10.1088/2041-8205/809/1/L14.

Barausse, E., Morozova, V. and Rezzolla, L. (2012). On the Mass Radiated by Coalescing Black Hole Binaries, *ApJ* **758**, 63, doi:10.1088/0004-637X/758/1/63.

Bardeen, J. M. and Petterson, J. A. (1975). The Lense-Thirring Effect and Accretion Disks around Kerr Black Holes, *ApJ* **195**, p. L65, doi:10.1086/181711.

Belczynski, K., Dominik, M., Bulik, T., O'Shaughnessy, R., Fryer, C. and Holz, D. E. (2010). The Effect of Metallicity on the Detection Prospects for Gravitational Waves, *ApJ* **715**, pp. L138–L141, doi:10.1088/2041-8205/715/2/L138.

Belczynski, K., Heger, A., Gladysz, W., Ruiter, A. J., Woosley, S., Wiktorowicz, G., Chen, H.-Y., Bulik, T., O'Shaughnessy, R., Holz, D. E., Fryer, C. L. and Berti, E. (2016a). The effect of pair-instability mass loss on black-hole mergers, *A&A* **594**, A97, doi:10.1051/0004-6361/201628980.

Belczynski, K., Holz, D. E., Bulik, T. and O'Shaughnessy, R. (2016b). The first gravitational-wave source from the isolated evolution of two stars in the 40-100 solar mass range, *Nature* **534**, pp. 512–515, doi:10.1038/nature18322.

Berti, E., Barausse, E., Cardoso, V., Gualtieri, L., Pani, P., Sperhake, U., Stein, L. C., Wex, N., Yagi, K., Baker, T., Burgess, C. P., Coelho, F. S., Doneva, D., De Felice, A., Ferreira, P. G., Freire, P. C. C. and Healy, J. et al. (2015). Testing general relativity with present and future astrophysical observations, *Classical and Quantum Gravity* **32**, 24, 243001, doi: 10.1088/0264-9381/32/24/243001.

Berti, E., Sesana, A., Barausse, E., Cardoso, V. and Belczynski, K. (2016). Spectroscopy of Kerr Black Holes with Earth- and Space-Based Interferometers, *Physical Review Letters* **117**, 10, 101102, doi:10.1103/PhysRevLett.117.101102.

Berti, E. and Volonteri, M. (2008). Cosmological Black Hole Spin Evolution by Mergers and Accretion, *ApJ* **684**, 822-828, doi:10.1086/590379.

Blanchet, L. (2014). Gravitational Radiation from Post-Newtonian Sources and Inspiralling Compact Binaries, *Living Reviews in Relativity* **17**, doi:10.12942/lrr-2014-2.

Bonetti, M., Haardt, F., Sesana, A. and Barausse, E. (2017). Post-Newtonian evolution of massive black hole triplets in galactic nuclei – II. Survey of the parameter space, *ArXiv e-prints 1709.06088* .

Buonanno, A. and Damour, T. (1999). Effective one-body approach to general relativistic two-body dynamics, *Phys. Rev. D* **59**, 8, 084006, doi:10.1103/PhysRevD.59.084006.

Buonanno, A. and Sathyaprakash, B. S. (2014). Sources of Gravitational Waves: Theory and Observations, *ArXiv e-prints 1410.7832* .

Callegari, S., Mayer, L., Kazantzidis, S., Colpi, M., Governato, F., Quinn, T. and Wadsley, J. (2009). Pairing of Supermassive Black Holes in Unequal-Mass Galaxy Mergers, *ApJ* **696**, pp. L89–L92, doi:10.1088/0004-637X/696/1/L89.

Cardoso, V. and Pani, P. (2017). Tests for the existence of black holes through gravitational wave echoes, *Nature Astronomy* **1**, pp. 586–591, doi:10.1038/s41550-017-0225-y.

Carr, B., Raidal, M., Tenkanen, T., Vaskonen, V. and Veermäe, H. (2017). Primordial black hole constraints for extended mass functions, *Phys. Rev. D* **96**, 2, 023514, doi:10.1103/PhysRevD.96.023514.

Colpi, M. (2014). Massive Binary Black Holes in Galactic Nuclei and Their Path to Coalescence, *Space Sci. Rev.* **183**, pp. 189–221, doi:10.1007/s11214-014-0067-1.

Colpi, M. and Sesana, A. (2017). *Gravitational Wave Sources in the Era of Multi-Band Gravitational Wave Astronomy*, pp. 43–140, World Scientific Publishing Co, doi:10.1142/9789813141766_0002.

Devecchi, B., Volonteri, M., Rossi, E. M., Colpi, M. and Portegies Zwart, S. (2012). High-redshift formation and evolution of central massive objects - II. The census of BH seeds, *MNRAS* **421**, pp. 1465–1475, doi:10.1111/j.1365-2966.2012.20406.x.

Dijkstra, M., Ferrara, A. and Mesinger, A. (2014). Feedback-regulated supermassive black hole seed formation, *MNRAS* **442**, pp. 2036–2047, doi:10.1093/mnras/stu1007.

Gair, J. R., Babak, S., Sesana, A., Amaro-Seoane, P., Barausse, E., Berry, C. P. L., Berti, E. and Sopuerta, C. (2017). Prospects for observing extreme-mass-ratio inspirals with LISA, in *Journal of Physics Conference Series*, *Journal of Physics Conference Series*, Vol. 840, p. 012021, doi:10.1088/1742-6596/840/1/012021.

Genzel, R., Eisenhauer, F. and Gillessen, S. (2010). The Galactic Center massive black hole and nuclear star cluster, *Reviews of Modern Physics* **82**, pp. 3121–3195, doi:10.1103/RevModPhys.82.3121.

Giacobbo, N., Mapelli, M. and Spera, M. (2018). Merging black hole binaries: the effects of progenitor's metallicity, mass-loss rate and Eddington factor, *MNRAS* **474**, pp. 2959–2974, doi:10.1093/mnras/stx2933.

Graham, A. W. and Spitler, L. R. (2009). Quantifying the coexistence of massive black holes and dense nuclear star clusters, *MNRAS* **397**, pp. 2148–2162, doi:10.1111/j.1365-2966.2009.15118.x.

Habouzit, M., Volonteri, M., Latif, M., Dubois, Y. and Peirani, S. (2016). On the number density of 'direct collapse' black hole seeds, *MNRAS* **463**, pp. 529–540, doi:10.1093/mnras/stw1924.

Haiman, Z., Kocsis, B. and Menou, K. (2009). The Population of Viscosity-

and Gravitational Wave-driven Supermassive Black Hole Binaries Among Luminous Active Galactic Nuclei, *ApJ* **700**, pp. 1952–1969, doi:10.1088/0004-637X/700/2/1952.

Haster, C.-J., Antonini, F., Kalogera, V. and Mandel, I. (2016). N-Body Dynamics of Intermediate Mass-ratio Inspirals in Star Clusters, *ApJ* **832**, 192, doi:10.3847/0004-637X/832/2/192.

Heger, A. and Woosley, S. E. (2002). The Nucleosynthetic Signature of Population III, *ApJ* **567**, pp. 532–543, doi:10.1086/338487.

Jiang, L., McGreer, I. D., Fan, X., Strauss, M. A., Bañados, E., Becker, R. H., Bian, F., Farnsworth, K., Shen, Y., Wang, F., Wang, R. and Wang, S. et al. (2016). The Final SDSS High-redshift Quasar Sample of 52 Quasars at z5.7, *ApJ* **833**, 222, doi:10.3847/1538-4357/833/2/222.

Khan, F. M., Fiacconi, D., Mayer, L., Berczik, P. and Just, A. (2016). Swift Coalescence of Supermassive Black Holes in Cosmological Mergers of Massive Galaxies, *ApJ* **828**, 73, doi:10.3847/0004-637X/828/2/73.

King, A. R., Lubow, S. H., Ogilvie, G. I. and Pringle, J. E. (2005). Aligning spinning black holes and accretion discs, *MNRAS* **363**, pp. 49–56, doi:10.1111/j.1365-2966.2005.09378.x.

Klein, A., Barausse, E., Sesana, A., Petiteau, A., Berti, E., Babak, S., Gair, J., Aoudia, S., Hinder, I., Ohme, F. and Wardell, B. (2016). Science with the space-based interferometer eLISA: Supermassive black hole binaries, *Phys. Rev. D* **93**, 2, 024003, doi:10.1103/PhysRevD.93.024003.

Kormendy, J. and Ho, L. C. (2013). Coevolution (Or Not) of Supermassive Black Holes and Host Galaxies, *ARA&A* **51**, pp. 511–653, doi:10.1146/annurev-astro-082708-101811.

Latif, M. A. and Ferrara, A. (2016). Formation of Supermassive Black Hole Seeds, *PASA* **33**, e051, doi:10.1017/pasa.2016.41.

Latif, M. A., Schleicher, D. R. G., Schmidt, W. and Niemeyer, J. C. (2013). The characteristic black hole mass resulting from direct collapse in the early Universe, *MNRAS* **436**, pp. 2989–2996, doi:10.1093/mnras/stt1786.

Lehner, L. and Pretorius, F. (2014). Numerical Relativity and Astrophysics, *ARA&A* **52**, pp. 661–694, doi:10.1146/annurev-astro-081913-040031.

Lupi, A., Colpi, M., Devecchi, B., Galanti, G. and Volonteri, M. (2014). Constraining the high-redshift formation of black hole seeds in nuclear star clusters with gas inflows, *MNRAS* **442**, pp. 3616–3626, doi:10.1093/mnras/stu1120.

Lupi, A., Haardt, F., Dotti, M., Fiacconi, D., Mayer, L. and Madau, P. (2016). Growing massive black holes through supercritical accretion of stellar-mass seeds, *MNRAS* **456**, pp. 2993–3003, doi:10.1093/mnras/stv2877.

Madau, P. and Dickinson, M. (2014). Cosmic Star-Formation History, *ARA&A* **52**, pp. 415–486, doi:10.1146/annurev-astro-081811-125615.

Madau, P. and Rees, M. J. (2001). Massive Black Holes as Population III Remnants, *ApJ* **551**, pp. L27–L30, doi:10.1086/319848.

Mandel, I., Brown, D. A., Gair, J. R. and Miller, M. C. (2008). Rates and Characteristics of Intermediate Mass Ratio Inspirals Detectable by Advanced LIGO, *ApJ* **681**, 1431-1447, doi:10.1086/588246.

Mapelli, M. (2016). Massive black hole binaries from runaway collisions: the impact of metallicity, *MNRAS* **459**, pp. 3432–3446, doi:10.1093/mnras/stw869.

Marconi, A., Risaliti, G., Gilli, R., Hunt, L. K., Maiolino, R. and Salvati, M. (2004). Local supermassive black holes, relics of active galactic nuclei and the X-ray background, *MNRAS* **351**, pp. 169–185, doi:10.1111/j.1365-2966.2004.07765.x.

Mayer, L., Fiacconi, D., Bonoli, S., Quinn, T., Rovskar, R., Shen, S. and Wadsley, J. (2015). Direct Formation of Supermassive Black Holes in Metal enriched Gas at the Heart of High-redshift Galaxy Mergers, *ApJ* **810**, 51, doi:10.1088/0004-637X/810/1/51.

Merloni, A. (2016). Observing Supermassive Black Holes Across Cosmic Time: From Phenomenology to Physics, in F. Haardt, V. Gorini, U. Moschella, A. Treves and M. Colpi. eds., *Lecture Notes in Physics, Berlin Springer Verlag*, *Lecture Notes in Physics, Berlin Springer Verlag*, Vol. 905, p. 101, doi:10.1007/978-3-319-19416-5_4.

Nandra, K., Barret, D., Barcons, X., Fabian, A., den Herder, J.-W., Piro, L., Watson, M., Adami, C., Aird, J., Afonso, J. M. and et al. (2013). The Hot and Energetic Universe: A White Paper presenting the science theme motivating the Athena+ mission, *ArXiv e-prints* .

Özel, F., Psaltis, D., Narayan, R. and McClintock, J. E. (2010). The Black Hole Mass Distribution in the Galaxy, *ApJ* **725**, pp. 1918–1927, doi:10.1088/0004-637X/725/2/1918.

Perego, A., Dotti, M., Colpi, M. and Volonteri, M. (2009). Mass and spin co-evolution during the alignment of a black hole in a warped accretion disc, *MNRAS* **399**, pp. 2249–2263, doi:10.1111/j.1365-2966.2009.15427.x.

Regan, J. A., Visbal, E., Wise, J. H., Haiman, Z., Johansson, P. H. and Bryan, G. L. (2017). Rapid formation of massive black holes in close proximity to embryonic protogalaxies, *Nature Astronomy* **1**, 0075, doi:10.1038/s41550-017-0075.

Reines, A. E., Greene, J. E. and Geha, M. (2013). Dwarf Galaxies with Optical Signatures of Active Massive Black Holes, *ApJ* **775**, 116, doi:10.1088/0004-637X/775/2/116.

Reisswig, C., Ott, C. D., Abdikamalov, E., Haas, R., Mösta, P. and Schnetter, E. (2013). Formation and Coalescence of Cosmological Supermassive-Black-Hole Binaries in Supermassive-Star Collapse, *Physical Review Letters* **111**, 15, 151101, doi:10.1103/PhysRevLett.111.151101.

Santamaría, L., Ohme, F., Ajith, P., Brügmann, B., Dorband, N., Hannam, M., Husa, S., Mösta, P., Pollney, D., Reisswig, C., Robinson, E. L., Seiler, J. and Krishnan, B. (2010). Matching post-Newtonian and numerical relativity waveforms: Systematic errors and a new phenomenological model for nonprecessing black hole binaries, *Phys. Rev. D* **82**, 6, 064016, doi:10.1103/PhysRevD.82.064016.

Sathyaprakash, B., Abernathy, M., Acernese, F., Ajith, P., Allen, B., Amaro-Seoane, P., Andersson, N., Aoudia, S., Arun, K., Astone, P. and et al. (2012). Scientific objectives of Einstein Telescope, *Classical and Quantum*

*Gravity* **29**, 12, 124013, doi:10.1088/0264-9381/29/12/124013.

Sathyaprakash, B. S. and Schutz, B. F. (2009). Physics, Astrophysics and Cosmology with Gravitational Waves, *Living Reviews in Relativity* **12**, doi:10.12942/lrr-2009-2.

Schleicher, D. R. G., Palla, F., Ferrara, A., Galli, D. and Latif, M. (2013). Massive black hole factories: Supermassive and quasi-star formation in primordial halos, *A&A* **558**, A59, doi:10.1051/0004-6361/201321949.

Schneider, R., Graziani, L., Marassi, S., Spera, M., Mapelli, M., Alparone, M. and Bennassuti, M. d. (2017). The formation and coalescence sites of the first gravitational wave events, *MNRAS* **471**, pp. L105–L109, doi:10.1093/mnrasl/slx118.

Schnittman, J. D. (2004). Spin-orbit resonance and the evolution of compact binary systems, *Phys. Rev. D* **70**, 12, 124020, doi:10.1103/PhysRevD.70.124020.

Sesana, A. (2016). Prospects for Multiband Gravitational-Wave Astronomy after GW150914, *Physical Review Letters* **116**, 23, 231102, doi:10.1103/PhysRevLett.116.231102.

Sesana, A., Barausse, E., Dotti, M. and Rossi, E. M. (2014). Linking the Spin Evolution of Massive Black Holes to Galaxy Kinematics, *ApJ* **794**, 104, doi:10.1088/0004-637X/794/2/104.

Sesana, A., Volonteri, M. and Haardt, F. (2007). The imprint of massive black hole formation models on the LISA data stream, *MNRAS* **377**, pp. 1711–1716, doi:10.1111/j.1365-2966.2007.11734.x.

Spera, M. and Mapelli, M. (2017). Very massive stars, pair-instability supernovae and intermediate-mass black holes with the SEVN code, *ArXiv e-prints* .

Spera, M., Mapelli, M. and Bressan, A. (2015). The mass spectrum of compact remnants from the PARSEC stellar evolution tracks, *MNRAS* **451**, pp. 4086–4103, doi:10.1093/mnras/stv1161.

Stone, N. C., Küpper, A. H. W. and Ostriker, J. P. (2017). Formation of massive black holes in galactic nuclei: runaway tidal encounters, *MNRAS* **467**, pp. 4180–4199, doi:10.1093/mnras/stx097.

Tamanini, N., Caprini, C., Barausse, E., Sesana, A., Klein, A. and Petiteau, A. (2016). Science with the space-based interferometer eLISA. III: probing the expansion of the universe using gravitational wave standard sirens, *JCAP* **4**, 002, doi:10.1088/1475-7516/2016/04/002.

Tremmel, M., Governato, F., Volonteri, M. and Quinn, T. R. (2015). Off the beaten path: a new approach to realistically model the orbital decay of supermassive black holes in galaxy formation simulations, *MNRAS* **451**, pp. 1868–1874, doi:10.1093/mnras/stv1060.

Valiante, R., Schneider, R., Graziani, L. and Zappacosta, L. (2018). Chasing the observational signatures of seed black holes at $z > 7$: candidate statistics, *MNRAS* **474**, pp. 3825–3834, doi:10.1093/mnras/stx3028.

van den Bosch, R. C. E. (2016). Unification of the fundamental plane and Super Massive Black Hole Masses, *ApJ* **831**, 134, doi:10.3847/0004-637X/831/2/134.

Vink, J. S. (2008). Mass loss and the evolution of massive stars, *New Astronomy*

*Reviews* **52**, pp. 419–422, doi:10.1016/j.newar.2008.06.008.

Volonteri, M. (2010). Formation of supermassive black holes, *A&A Rev.* **18**, pp. 279–315, doi:10.1007/s00159-010-0029-x.

Woosley, S. E. (2017). Pulsational Pair-instability Supernovae, *ApJ* **836**, 244, doi:10.3847/1538-4357/836/2/244.

Wu, X.-B., Wang, F., Fan, X., Yi, W., Zuo, W., Bian, F., Jiang, L., McGreer, I. D., Wang, R., Yang, J., Yang, Q., Thompson, D. and Beletsky, Y. (2015). An ultraluminous quasar with a twelve-billion-solar-mass black hole at redshift 6.30, *Nature* **518**, pp. 512–515, doi:10.1038/nature14241.

Zlochower, Y., Campanelli, M. and Lousto, C. O. (2011). Modeling gravitational recoil from black-hole binaries using numerical relativity, *Classical and Quantum Gravity* **28**, 11, 114015, doi:10.1088/0264-9381/28/11/114015.

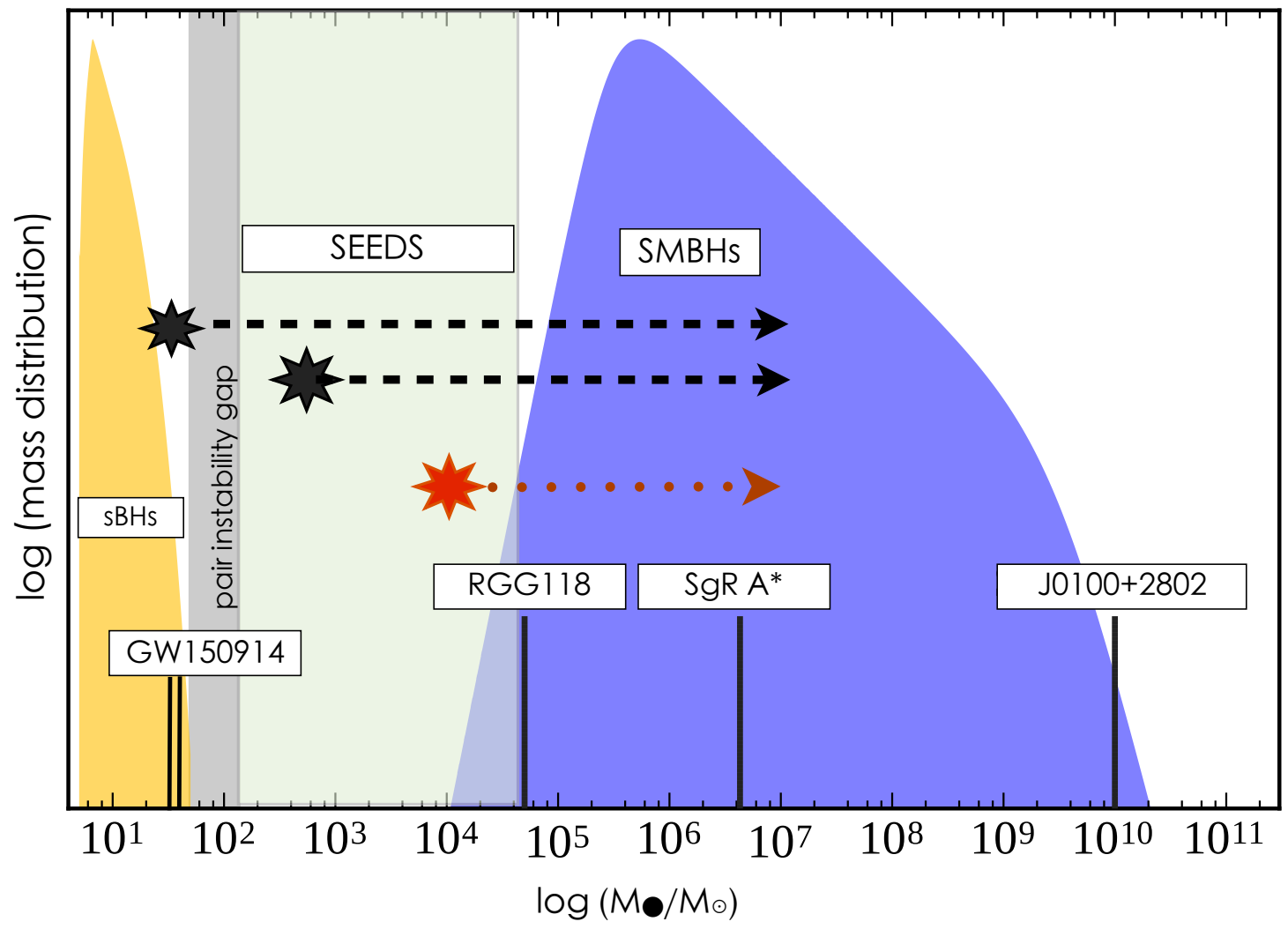

Fig. 1. Cartoon illustrating the BH mass spectrum encompassing the whole astrophysical relevant range, from sBHs to SMBHs, through the unexplored (light-green) zone where seed BHs are expected to form and grow. Vertical black-lines denote the two sBH masses in GW150914 (Abbott *et al.*, 2016a), the mass $M_\bullet$ of RGG118 (the lightest SMBH known as of today in the dwarf galaxy RG118), of SgrA* in the Milky Way (Genzel *et al.*, 2010), and of J0100+2802 (the heaviest SMBH ever recorded). The mass distribution of sBHs, drawn from the observations of the Galactic sBH candidates, has been extended to account for the high-mass tail following the discovery of GW150914, which host sBHs of $\sim 30\,\mathrm{M}_\odot$, in the binary prior merging. The minimum (maximum) sBHs is arbitrarily set equal to $3\,\mathrm{M}_\odot$ ($60\,\mathrm{M}_\odot$), and the theoretically predicted pair-instability gap is depicted as narrow darker-grey strip. The SMBH distribution has been drawn scaling their mass according to the local galaxy mass function and $M_\bullet$-$\sigma$ correlation. The decline below $\sim 10^5\,\mathrm{M}_\odot$ is set arbitrary: BH of $\sim 10^{4-5}\,\mathrm{M}_\odot$ may not be ubiquitous in low-mass galaxies as often a nuclear star cluster is in place in these galaxies, which may or may not host a central IMBH (Graham and Spitler, 2009). The black stars and dashed tracks illustrate the possibility that a SMBH forms as sBH-only (born on the left side of the sBH gap) or as light seed (on the right of the gap) which then grows through phases of super-Eddington accretion (Lupi *et al.*, 2016). The red star and dotted track illustrates the possibility of a *genetic* divide between sBHs and SMBHs, and that a heavy seed forms through the direct collapse of a supermassive isolated cloud/protostar in a metal free, atomic-hydrogen cooling, dark-matter halo (Latif *et al.*, 2013; Schleicher *et al.*, 2013). The seed later grows via gas accretion and mergers with other halos. LIGO-Virgo, the third generation of Ground Based Interferometers, and the LISA space mission will shed light into the physical mechanisms leading to the formation of SMBHs in the unexplored range between sBHs and SMBHs.

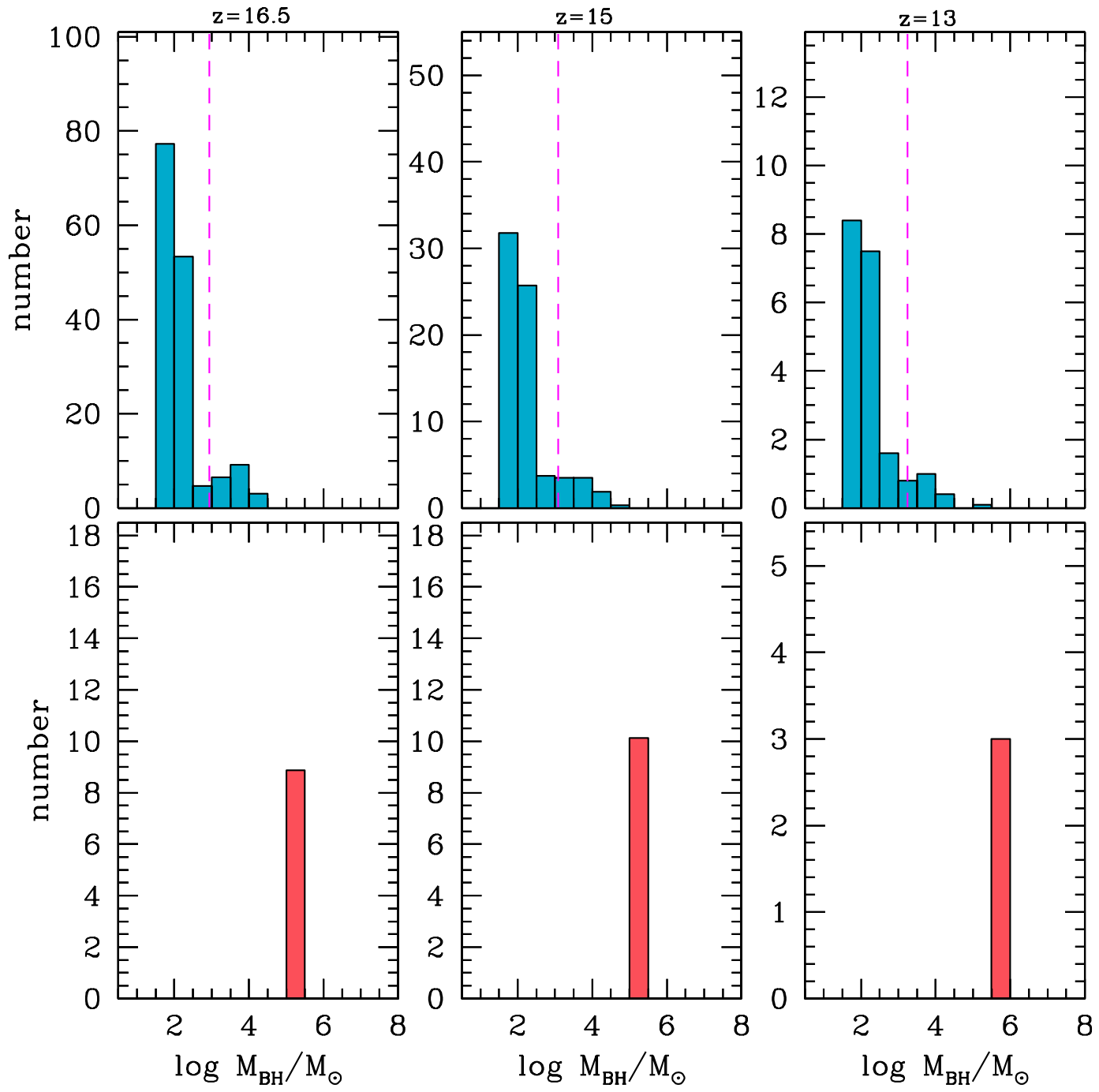

Fig. 2. Mass distribution of *light seeds* (top panel) and *heavy seeds* (bottom panel) born in dark matter halos of $\sim 10^8$ $M_{\odot}$ at $z > 16.5$, at three different redshifts $z = 16.5, 15, 13$, in a biased region of the universe selected to explain the existence of SMBHs at $z \sim 7$. The seeds live in isolated halos for about 200 Myrs, before halo-halo mergers occur. During this lapse time, heavy seeds grow faster than light seeds that do not modify significantly their mass spectrum. The dashed lines in the upper panels indicate the average mass of isolated light seed at the corresponding $z$. Courtesy of R. Valiante and R. Schneider (Valiante *et al.*, 2018).

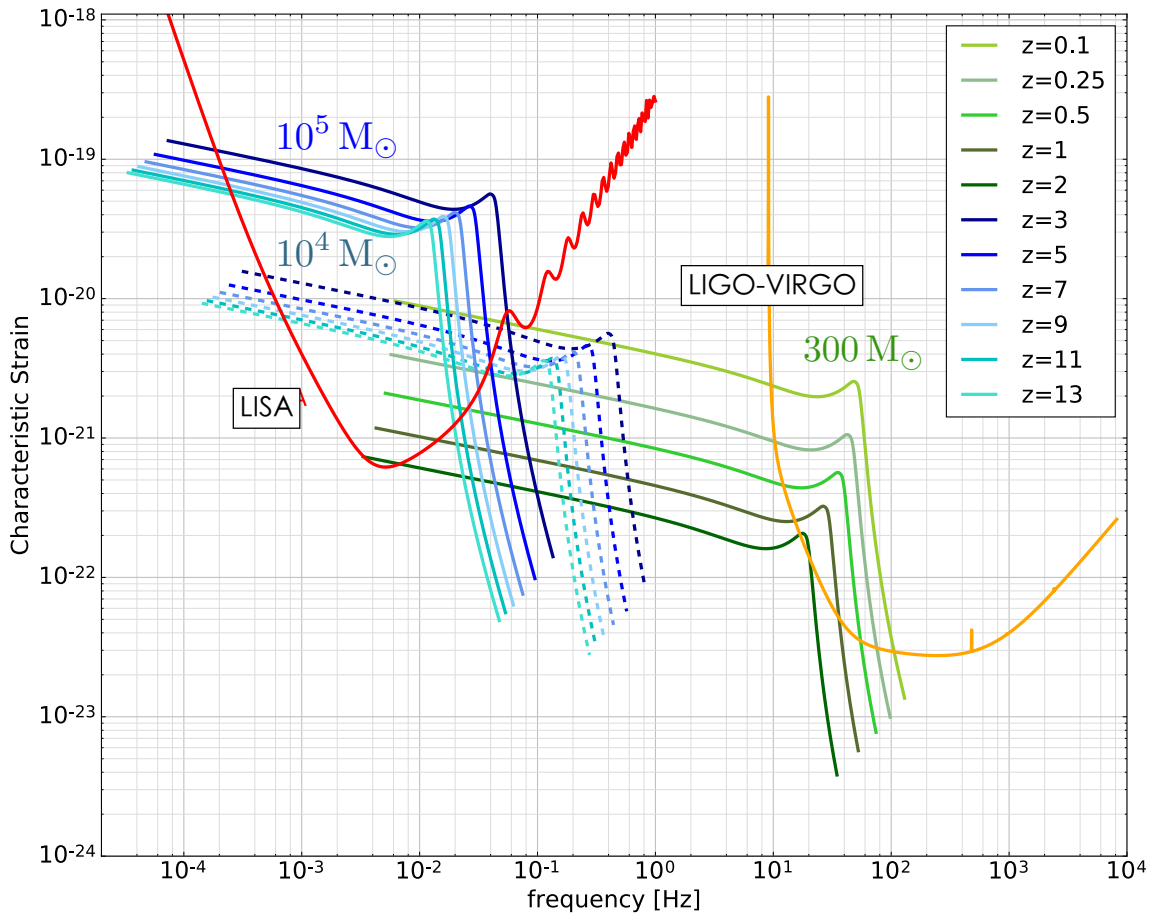

Fig. 3. Red curve depicts the sensitivity curve of the LISA observatory in its current design (Amaro-Seoane *et al.*, 2017), expressed in terms of the strain noise $(fS_{\rm noise})^{1/2}$ as a function of the observer-frame frequency $f$. The GW signal from coalescing seed BHs in circular binaries at selected redshifts is described in terms of the sky, inclination and polarization averaged characteristic strain $f\tilde{h}(f)$ (in dimensionless units) inferred using non-precessing waveform models (Santamaría *et al.*, 2010): upper solid and dashed lines in blu and light blu colors refer to signals form BBHs with mass ratio $q = 1$ and total rest-frame mass $M_{\rm B} = 10^5\,{\rm M}_\odot$ and $10^4\,{\rm M}_\odot$, respectively. The two family of lines refer to BH binaries detected at different redshift, as indicated in the inset, from $z = 3$ up to $z = 13$. Lower yellow and dark-green solid curves refer to IMBH binaries with total mass $M_{\rm B} = 300\,{\rm M}_\odot$ and $q = 1$; colors are associated to different redshifts varying between 0.1 and 2, as indicated in the inset with corresponding colors. Notice that for each BBH of given mass $M_{\rm B}$, the frequency at which $f\tilde{h}(f)$ is maximum (i.e. near coalescence) shifts to lower values with increasing redshift due to cosmic expansion. Seeds of $10^4\,{\rm M}_\odot$ do not coalesce in the LISA band, and can be detected during their inspiral phase only, out to very large redshifts. IMBH binaries with mass of $300\,{\rm M}_\odot$ are detectable first in the LISA band, during their slow adiabatic contraction, and later at the time of coalescence in the LIGO-Virgo band, only if they are at low redshift $z \lesssim 1$ (as also described in Figure 4).

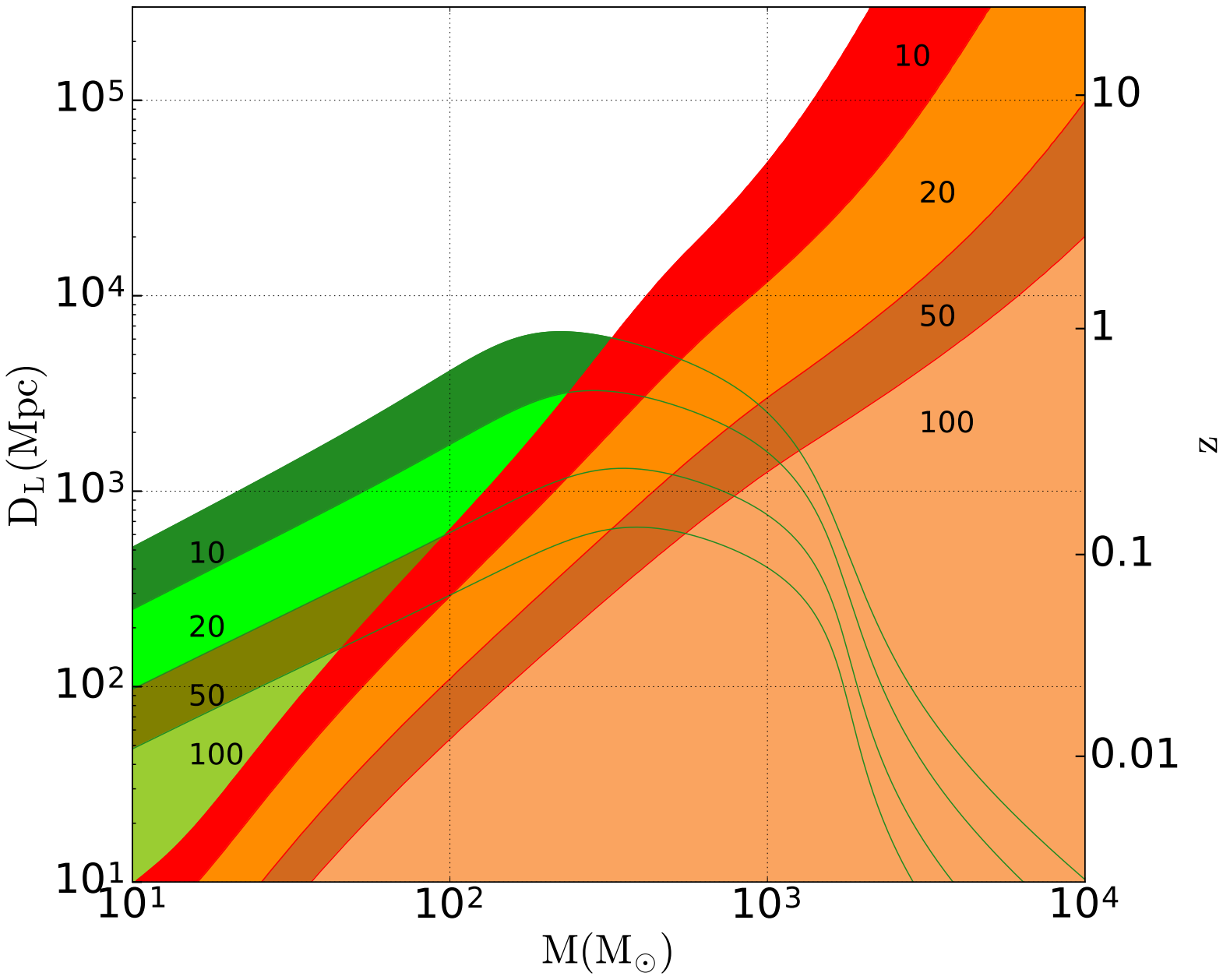

Fig. 4. $D_L - M$ plane, where on the left $y$-axis $D_L$ is the luminosity distance (in units of Mpc), on the right $y-$axis the redshift $z$, and on the $x-$axis the total mass of the binary $M_{\rm B}$ in the the source-frame. Lines of constant sky, inclination and polarization averaged SNR are drawn for non-spinning, equal mass binaries hosting sBHs and IMBHs. The SNR is computed using the waveform models of Santamaría *et al.* (2010). The figure illustrates the synergy between LISA and advanced LIGO in the common effort to search for low redshift sBH and IMBH binaries. LISA detects the sources in the inspiral phase, whereas LIGO in the merger proper.

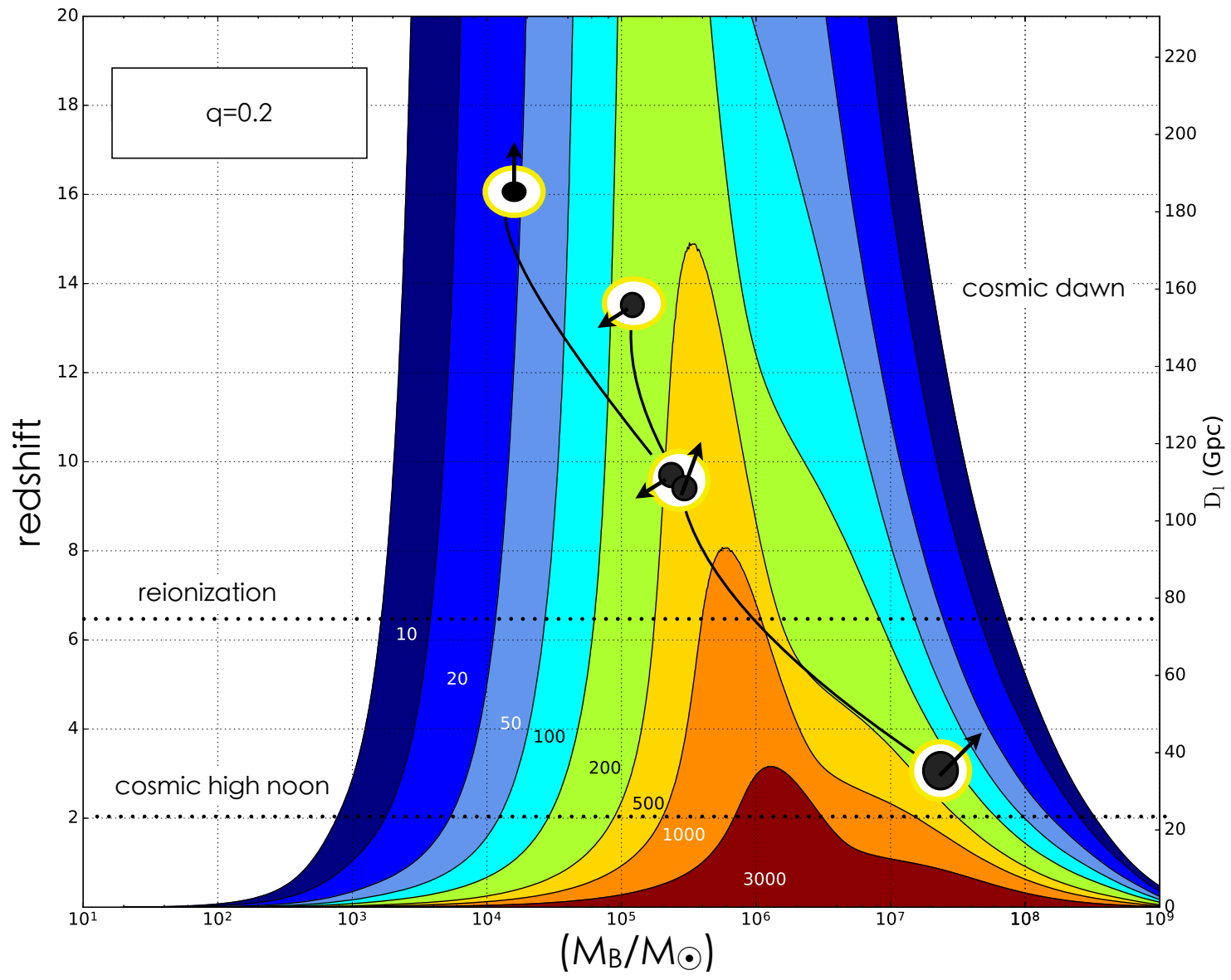

Fig. 5. The $z - M_{\rm B}$ plane, where $z$ is the redshift and $M_{\rm B}$ the rest-frame BBH total mass in units of $M_\odot$. The luminosity distance $D_{\rm l}$ is plotted on the right of the $y$-axis. Lines of constant (sky-polarization-and-inclination averaged) SNR are drawn for spinning (non-precessing) BBHs, with mass ratio $q = 0.2$. The spins are aligned with the orbital angular momentum and have equal magnitude $\chi_{\rm spin} = 0.9$. The SNR is computed using the waveform models of Santamaría *et al.* (2010). Dotted lines refer to the redshifts of: *cosmic high noon* ($z \sim 2$, corresponding to the peak of the averaged star formation rate and AGN activity); *cosmic reionization* at about $z \sim 6$, corresponding to the end of epoch when interstellar hydrogen transited from being neutral to being ionized); and *cosmic dawn* (between $20 < z < 9$, where the first luminous objects star to form). We inserted a BH evolutionary track to show that BHs inevitably cross the LISA bandwidth. Black dots represent BHs, the arrows the spin vectors **S**. BHs are embedded in galactic halos (white-yellow circles) and experience episodes of accretion (black lines) and coalescences inside merged halos, which increase the mass and modify the spin. The selected track illustrates the formation of a heavy BH active during cosmic high noon.

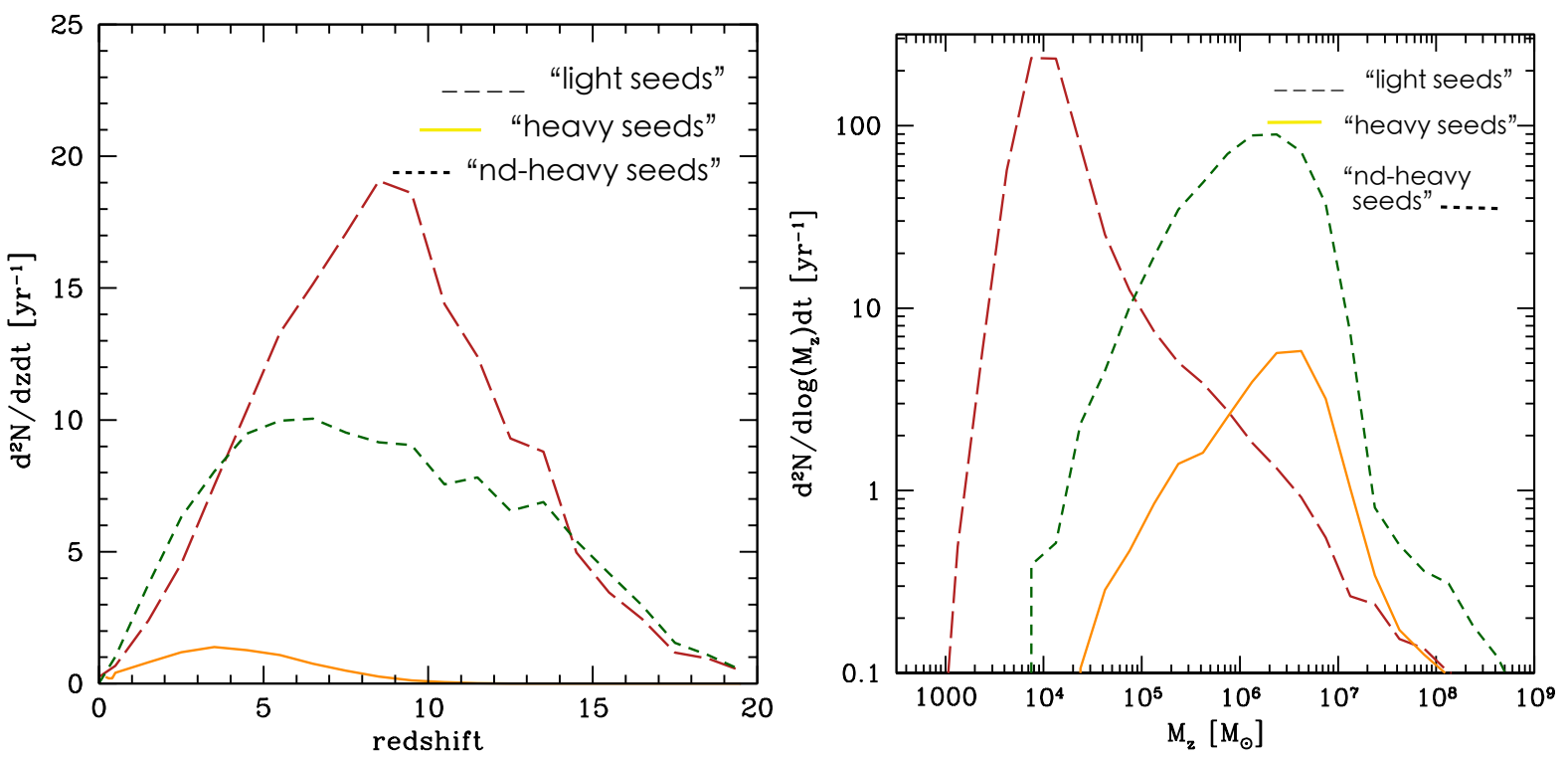

Fig. 6. Predicted BBH coalescence rates per unit redshift (left panel) and per unit binary mass (observer-frame) $M_z = (1+z)M_{\rm B}$ (right panel) for three seed models, from Klein et al. (Klein *et al.*, 2016). The semi-analytical model used follows the evolution of baryonic structures along a dark-matter merger tree produced by an extended Press-Schechter formalism, modified to reproduce results of N-body simulations[Klein *et al.* (2016)]. In the light-seed scenario, seeds are drawn from a log-normal distribution for the progenitor stars centered around $300\,{\rm M}_\odot$ with a rms of 0.2 dex and the exclusion region between $140-260\,{\rm M}_\odot$ corresponding to the onset of pair instabilities. Seed BHs carry a mass equal to 2/3 of the mass of the progenitor star. The seeds are implanted in rare massive halos collapsing from the $3.5\sigma$ peaks of the primordial density field and are allowed to accrete at twice the Eddington limit. In the heavy-seed scenario, seed BHs carry masses of $10^5\,{\rm M}_\odot$ at form before redshift $z\sim 15$. Both light and heavy seed models account for delays between the formation of the binary in a halo-halo merger and the time of its merging. GW driven coalescence acts on very small galactic scales and dissipative mechanisms driven by the interaction of stars or gas with the binary set the conditions for its coalescence on timescale less than the current age of the universe. Delays have been distributed over a range that varies according to the gas or/and stellar content [rapid (long) when gas (stars) is (are) present, i.e. of order $10^8$ yr ($10^9$ yr)] (see Antonini *et al.* (2015) for details). The figure illustrates how sensitive are the merger rates to the physical models for the seeds and delay time distributions. Courtesy of A. Klein (Klein *et al.*, 2016).